\documentclass[hidelinks,journal]{IEEEtran}
\usepackage{url,array,lipsum,hhline,graphicx,subfigure,amsfonts,cite,epstopdf,makeidx,outlines,textcomp,multirow,soul,hyperref,enumitem,amssymb,amsmath,color,graphicx, amsbsy}
\usepackage[table,xcdraw]{xcolor}
\usepackage[linesnumbered,ruled]{algorithm2e}
\usepackage[noprefix]{nomencl}
\usepackage[none]{hyphenat}
\allowdisplaybreaks
\usepackage{clipboard}
\usepackage{makecell}

\begin{document}
\bstctlcite{IEEEexample:BSTcontrol}
\title{Natural Gas Short-Term Operation Problem with Dynamics: A Rank Minimization Approach }

\author{ {Reza Bayani, \emph{Student Member, IEEE} and Saeed D. Manshadi, \emph{Member,  IEEE}}
\thanks{Reza Bayani is with University of California San Diego, La Jolla, CA, 92093, USA, and San Diego State University, San Diego, CA, 92182, USA. Saeed D. Manshadi is with San Diego State University, San Diego, CA, 92182, USA. e-mail: rbayani@ucsd.edu; smanshadi@sdsu.edu.}
\vspace{-.5cm}
}

\markboth{
}%
{Shell \MakeLowercase{\textit{et al.}}: Bare Demo of IEEEtran.cls for Journals}

\maketitle

\begin{abstract}
Natural gas-fired generation units can hedge against the volatility in the uncertain renewable generation, which may occur during very short periods.  It is crucial to utilize models  capable of correctly capturing the natural gas network dynamics induced by the volatile demand of gas-fired units. The Weymouth equation is commonly implemented in literature to avoid dealing with the mathematical complications of solving the original governing differential equations of the natural gas dynamics.   However, it is shown in this paper that this approach is not reliable \emph{in the short-term operation problem}. Here, the merit of the non-convex transient model is compared with the simplified Weymouth equation, and the drawbacks of employing the Weymouth equation are illustrated. The results demonstrate how changes in the natural gas demand are met by adjustment in the pressure within pipelines rather than the output of natural gas suppliers. This work presents a convex relaxation scheme for the original non-linear and non-convex natural gas flow equations with dynamics, utilizing a rank minimization approach to ensure the tightness. The proposed method renders a computationally efficient framework that can accurately solve the non-convex non-linear gas operation problem and accurately capture its dynamics. Also, the results suggest that the proposed model improves the solution optimality and solution time compared to the original non-linear non-convex model. Finally, the scalability of the proposed approach is verified in the case study. 

\end{abstract}
\begin{IEEEkeywords}
natural gas dynamics, convex relaxation, rank minimization, uncertainty, short-term operation. \vspace{-0.2cm}
\end{IEEEkeywords}
\IEEEpeerreviewmaketitle 
\section*{Nomenclature} \vspace{-.5cm}
\subsection*{Variables}
\noindent \begin{tabular}{ l p{6.55cm} }
$d_{u,t}$ & Served gas demand of gas-fired unit\\
$m^t_{p,s}$ & Mass flow rate through segments of pipes\\
$pr^{\textbf{(.)}}$ & Pressure at junctions or segments of pipe ($Pa$)\\
$P^{\textbf{(.)}}$ & Real power dispatch\\
$v_{g,t}^G$ & Gas supply of supplier $g$ at time $t$ $(kcf)$\\
$\gamma^t_{p,s}$ & Lifting variable associated with pipe segments\\
$\theta_{b,h}$ & Voltage angle of bus $b$ at hour $h$\\
$\eta^{\textbf{(.)}}_{\textbf{(.)}},\zeta^{\textbf{(.)}}_{\textbf{(.)}},\lambda^{\textbf{(.)}}_{\textbf{(.)}}$ & Dual variables for equality constraints\\
$\underline{\mu}^{\textbf{(.)}}_{\textbf{(.)}},\overline{\mu}^{\textbf{(.)}}_{\textbf{(.)}}$ & Dual variables for inequality constraints\\
\end{tabular}\vspace{-.15cm}
\subsection*{Sets}\vspace{-0cm}
\noindent \begin{tabular}{ l p{7.2cm} }
$b \in \mathcal{B}$  & Set of buses in power network\\
$br \in \mathcal{BR}$  & Set of branches in power network\\
$c \in \mathcal{C}$  & Set of junctions with gas compressors\\
$g \in \mathcal{G}$  & Set of gas suppliers\\
$h \in \mathcal{H}$  & Set of hours\\
$i \in \mathcal{I}$  & Set of units in power network\\
$p \in \mathcal{P}$  & Set of pipes in gas network\\
$s \in \mathcal{S}_p$  & Set of all segments of pipe $p$\\
$t \in \mathcal{T}$  & Set of time intervals\\
$u \in \mathcal{U}$  & Set of gas-fired units in the power network\\
\end{tabular}
\subsection*{Parameters} \vspace{-0cm}
\noindent \begin{tabular}{ l p{7.2cm} }
$c$ & Speed of sound $(m/s)$\\ 
$D$ & Diameter of natural gas pipe  $(m)$\\
$d^G_{l,t}$ & Demand of natural gas load $l$ at time $t$\\
$f$ & Friction factor of natural gas pipes\\
$F^C_i,F^G_u$ & Cost function/ Gas consumption function of unit $i$\\
$\underline{pr}_j^{J},\overline{pr}_j^{J}$ & Lower/upper bounds for pressure {at} junction $j$\\ 
$\overline{P}^{D}_{b,h}$ & Real power demand of bus $b$ at hour $h$\\
$\underline{v}_g^{G},\overline{v}_g^{G}$ & Lower/upper bounds for natural gas supplier $g$\\
$x_{br}$ & Reactance of branch $br$\\
$\Gamma$ & Ratio of compressors in natural gas network\\
$\Delta x$ & Length of natural gas pipe segments\\
$\Delta t$ & Time step duration\\
$\kappa_{E},\kappa_{G}$ & Value of lost load in electricity/gas networks\\
$\xi_g$ & Cost multiplier of natural gas supplier $g$ $(\$/kcf)$\\
\end{tabular}
\section{Introduction} 
\begin{figure}[h!]\vspace{-.35cm}\centering
{\includegraphics[width=.85\columnwidth]{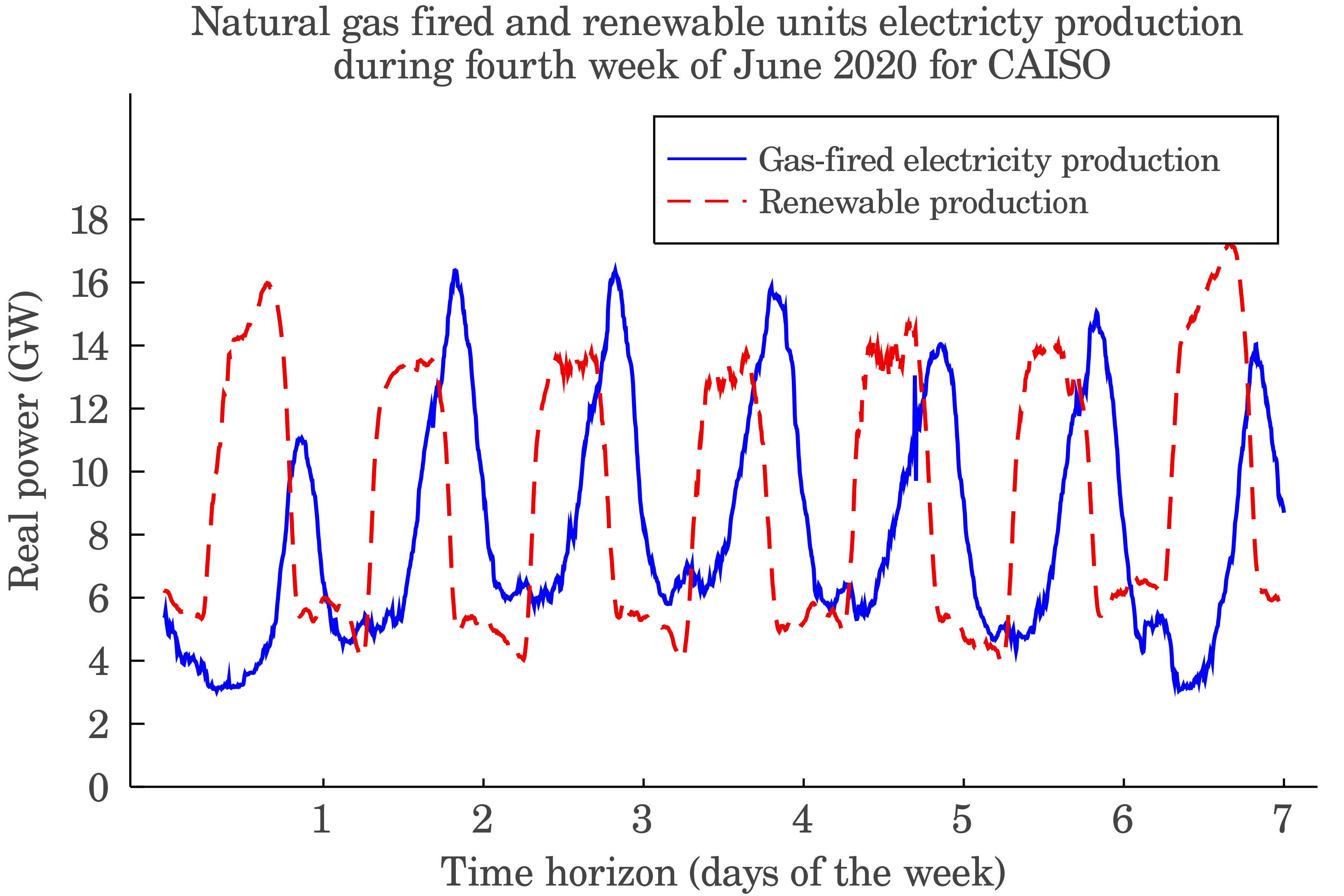}}
\vspace{-.35cm}
\caption{Gas based and renewable electricity generation 22-28 June 2020 }
\label{fig:ng_production}
\vspace{-.55cm}
\end{figure}

\IEEEPARstart{C}URRENTLY, natural gas-fired generation accounts for the largest portion of the electricity generation among all types of electricity production in the United States \cite{eia2020}. It is estimated that the share of natural gas-fired units will remain at 37\% among all types of electricity generation in the US for the next three decades. Due to their fast-responsive nature, natural gas-fired generation units are mostly utilized during peak hours. In addition to meeting demand fluctuations, they also provide support when renewable generation falls short of the predicted values. The rotatory outage that occurred during the 2021 winter storm of Texas is an example of a renewable scarcity event. Besides, in a highly renewable integrated network {such as} California's, the dispatch of natural gas-fired units is doubled or tripled in {the span of a }few hours several days a year. According to the data provided by California Independent System Operator (CAISO), it is observed from Fig. \ref{fig:ng_production} that natural gas-based generations experience evening surges every day \cite{caisowebsite}. These patterns are mostly caused by the simultaneous drop in total solar and wind generation {supply} and the diurnal growth in electricity consumption. It is noticed from Fig. \ref{fig:ng_production} that the amount of variations in daily renewable power generation is in the order of GW per hour. Currently, natural gas-fired units are deemed the practical option to hedge against uncertainties of this magnitude due to their extensive penetration.

\par The operation problem of joint electricity and natural gas networks has been a subject of many researchers \cite{zhang2015hourly,alabdulwahab2015coordination,he2016robust,qiu2015linear,ding2017multi,qadrdan2013operating,correa2014integrated,zlotnik2016coordinated,manshadi2018coordinated}. Among the applications studied in this area, we can mention demand response \cite{zhang2015hourly}, renewable generation integration \cite{alabdulwahab2015coordination}, expansion planning \cite{he2016robust,qiu2015linear}, and market operations \cite{ding2017multi}. Natural gas linepack modeling and its effects are also covered in several research works \cite{qadrdan2013operating,correa2014integrated,he2016robust}. The extent of coordination between electricity and natural gas networks is an important aspect. In some cases, natural gas network equations are merely considered as constraints (i.e., fuel security-constrained) for the optimization problem of the electricity network \cite{correa2014integrated}. On the contrary, sometimes it is the case that the operation of these networks is fully coordinated, and each network is aware of the other network's state \cite{zlotnik2016coordinated,manshadi2018coordinated}. In any case, regardless of the application or the degree of coordination, utilizing a model capable of correctly capturing and representing the natural gas network dynamics is critical.

\par Natural gas demand is highly varying and depends on various elements including the changes in the electricity demand and non-dispatchable renewable generation. Since natural gas variations influence electricity operations, precise modeling of gas transients is essential to the coupled operation of electricity and natural gas infrastructures. The choice of natural gas model is often a trade-off between solution time and accuracy. Naturally, models with more accurate representations of natural gas dynamics come at the price of higher complexities and subsequent computational challenges. In some studies, researchers might not even consider any gas flow model \cite{wu2011optimal,zhao2016unit,conejo2020operations}. These works usually deal with midterm and long-term planning horizons, and understandably, neglect gas flow modeling. {However, in short-term planning horizons, the impact of variations distinctly stands out and natural gas flow \emph{dynamics} cannot be disregarded. In short-term problems, the choice of the OGF model becomes exceedingly important since the available time window for running the model is limited.} From the natural gas dynamics point of view, the existing research work coping with the coordinated scheduling of electricity and natural gas networks is best categorized as \textit{steady-state} and \textit{transient} models.

\vspace{-.3cm}\subsection{Literature Review on Steady-State Models}
To avoid the computational burden of dealing with the differential equations of the Optimal Gas Flow (OGF) problem, simplified steady-state models are {often} used in the literature. The general flow equation, which considers variations in the length of the pipe and its diameter along with certain assumptions, can be approximated to the widely applied Weymouth equation \cite{geidl2007optimal,liu2011coordinated,qadrdan2013operating,alabdulwahab2015coordination,ding2017multi}. However, as will be shown later in this paper, the \emph{Weymouth equation is unsuitable for short-term operation problems}.  In some research works, the authors have approximated the Weymouth model with linearization techniques \cite{correa2014integrated,zhang2015hourly,manshadi2015resilient}. 
Another approach is opting for convex relaxation of the Weymouth formulation \cite{he2016robust,manshadi2018tight,qiu2015linear,manshadi2018coordinated,jia2020convex}, with Second-Order Cone Programming (SOCP) \cite{he2017decentralized,fan2020multi,Chen2018} and Semi-definite Programming (SDP) \cite{ojha2017solving} being among the utilized methods.  

\par Steady-state approximations lead to inaccurate results in short-term operation problems which are subject to highly varying natural gas consumption rates. These approaches neglect dynamic relations between natural gas pressure and mass flow rate, caused by gas compressibility and low velocity, both inside pipelines and across different time steps. That is why for short-term planning periods, it is better to avoid using steady-state models and {instead} implement transient models which accurately capture the intra-hour dynamics of natural gas through the spatio-temporal representation of the system. Besides, steady-state models assume that {the} supply-demand balance in a natural gas network is established momentarily. In reality, it could take several hours for natural gas to travel from the source to the demand junction, depending on the distance. Throughout this article, the phrase ``short-term operation"  is best categorized as operation planning problems in the span of {a} few hours up to a day.

\vspace{-.35cm}\subsection{Literature Review on Transient Models}
Transient models obtain the OGF problem by considering the governing Partial Differential Equations (PDEs) of the natural gas network \cite{zlotnik2015model,zlotnik2016coordinated,badakhshan2019security}. Finite diﬀerence methods are implemented to transform these PDEs into time-diﬀerence equations so that the governing PDEs of the OGF problem can be numerically solved \cite{liu2019optimal}. However, finding the optimal solution to this non-linear, non-convex, and spatio-temporally discretized optimization problems is still challenging  in terms of calculation time. In situations with shorter planning  periods, the extended solution time required for solving these models may not render them practical for such applications. Additionally, often these models entail utilizing non-linear solvers, which do not guarantee convergence to the optimal solution. Linearization approaches are frequently adopted to negate the inherent non-linearity of these equations and reduce solver time \cite{fang2017dynamic,yang2017effect,qi2019decentralized,zhou2017equivalent}. Although linearization techniques are a way out of the extremely high computation burden caused by transient OGF models, they still rely on consequential approximations. Besides, they are not mathematically capable of representing the exact dynamics of natural gas. Additionally, linearization methods are problem-specific and cannot be generalized into all tasks since they are applied around setpoints. 

\par
\begin{figure}[t]
	\vspace{-1.15cm}
    \centering
    \includegraphics[width=.8\linewidth]{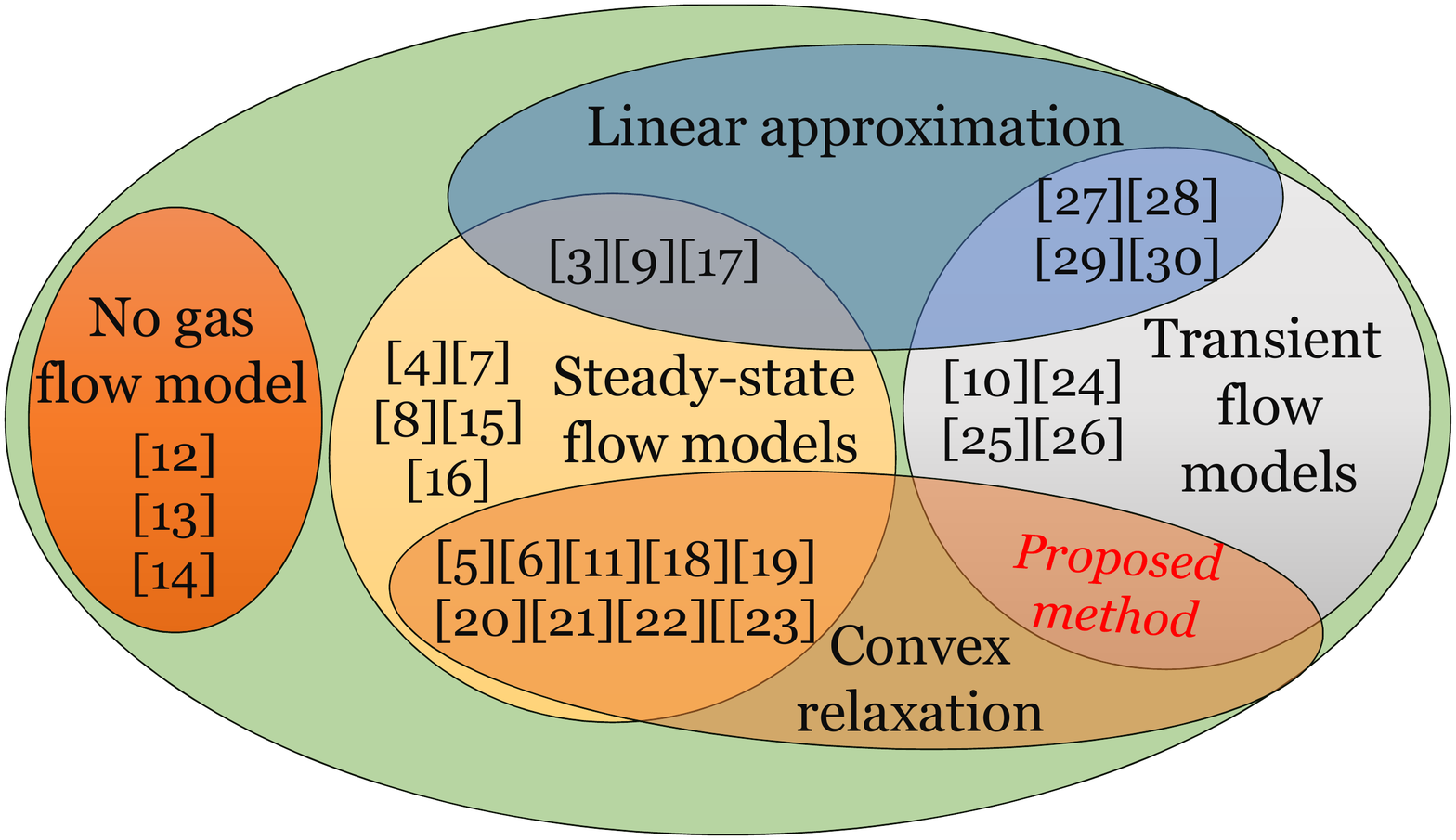}
    \vspace{-0.95cm}
    \caption{Graphical representation of the modeling approaches in the literature}
    \vspace{-0.45cm}
  \label{fig:ven_review}
  \vspace{-.15cm}
\end{figure}

 Fig. \ref{fig:ven_review} illustrates the current state of the literature dealing with the OGF problem. This work leverages the discretized transient flow problem formulation, which is solved via the proposed tight relaxation scheme. 
According to the literature, no work has dealt with convex relaxation of the non-linear transient OGF problem. The present article aims to fill this gap by proposing a convex relaxation method by introducing a rank minimization technique using a bi-level optimization formulation. Avoiding approximate methods enables us to enjoy the exactness and accuracy offered by the transient flow models. Inaccuracies in the procured solution to the OGF problem could evolve into potentially costly underestimation or overestimation of the natural gas network and its interactions with gas-fired units. 

\par Moreover, the major issue with non-linear transient models is their inevitable time-consuming computations, which grow exponentially with the size of the problem. The proposed relaxation method drastically reduces the solver time compared with the non-linear and non-convex models. Finally, utilizing non-linear solvers does not guarantee the optimal solution, whereas convex relaxation allows reaching an optimal solution within definite relaxation gaps. Although steady-state models are solved quickly, they utilize simplifying assumptions that {are} appropriate only for long-term planning horizons. Since transient models utilize fewer approximations and consider smaller time intervals, they are suitable for short-term operation periods. However, the major downside of utilizing high-granularity transient OGF models is their extensive computation burden. This work aims to solve this issue by proposing a tight convex relaxation for the transient OGF model. Objectively, any entity dealing with the operation of natural gas-fired generation units can benefit from the proposed model. These entities include ISOs, regional transmission organizations, generation companies, and gas companies.

\vspace{-.35cm}

\subsection{Contributions}
The contributions of this work are listed as follows:
\begin{enumerate}[wide, labelwidth=!,labelindent=10pt]
    \item The limitations of the Weymouth equation, which presents an approximation of natural gas behavior, in tackling high fluctuations in renewable generation in interdependent electricity and natural gas networks are illustrated. We demonstrate how leveraging the transient model ensures no errors will arise in modeling natural gas behavior in response to sudden changes in demand. At the same time, the Weymouth equation fails to present a reliable model for fuel availability.
    \item The presented convex relaxation scheme for the non-linear non-convex transient OGF problem is tight, i.e, the solution rendered by the relaxed problem is feasible for the original non-convex problem. A bi-level problem is formed to implement the proposed rank minimization technique. The presented form is superior to the standard rank minimization procedure where the objective of the relaxed OGF problem is penalized with a perturbation term. This approach distinguishes between rank minimization in the upper-level and cost minimization in the lower-level objective functions. As a result, a near-rank-1 solution is obtained, which is also optimal at the same time. 
    \item A computationally advantageous scheme for solving the OGF problem is obtained as a result of employing the proposed convexification and rank minimization framework. A low solution time makes it a perfect tool for short-term operation problems. This fact is illustrated by comparing the solution time of the proposed model with {that of} the original non-convex model. Utilizing the proposed convex model renders a physically meaningful optimal solution with better objective value and a solution time in orders of magnitude faster than the original non-linear and non-convex model.
    \item Several key findings are illustrated, including the fact that the ﬂuctuations in the natural gas pipelines will address the changes in the natural gas demand rather than the changes in the dispatch of resources. This modeling perspective is crucial for the short-term operation and the preparedness of the natural gas network to meet the volatile natural gas demand of gas-fired units. 
    \end{enumerate}
\vspace{-.35cm}
\section{Problem Formulation}
In this paper, renewable integrated electricity grids which contain gas-fired generators are considered. The power network is coupled with a natural gas network, so the production level of the electricity grid is dependent on the natural gas provision. While the operation problem presented here focuses on the look-ahead periods in the order of few hours, the given problem can also be generalized for the day-ahead operation problem.
\vspace{.15cm}
\subsubsection{Electricity Network Look-Ahead Operation Problem}
The look-ahead operation problem for the electricity network is a short-term operation problem as shown in \eqref{eq:elec_side}. \vspace{-.15cm}
\begin{subequations} \label{eq:elec_side}
\begin{alignat}{3}
&\underset{P}{\min} \sum_{h\in\mathcal{H}}\sum_{i\in\mathcal{I}}^{}{F^C_i({P}^{Gen}_{i,h})}+\kappa_{E}\sum_{h\in\mathcal{H}}\sum_{b\in\mathcal{B}}^{}{(\overline{P}^{D}_{b,h} -P^{D}_{b,h})} \label{eq:elec_obj}\\
&\textit{subject to:} \nonumber \\
& P^{Br}_{br,h}\cdot{x_{br}} = {\theta_{{b}^{fr}_{br},t}-\theta_{{b}^{to}_{br},t}}, \hspace{1.1cm} \forall\;br\in \mathcal{BR}, h\in\mathcal{H}\label{eq:elec_dc}\\
& \underline{P}^{Gen}_{i} \leq {P}^{Gen}_{i,h} \leq \overline{P}^{Gen}_{i},\hspace{2.3cm}\forall i\in \mathcal{I}, h\in\mathcal{H}\label{eq:elec_p_limit}\\
& |P^{Br}_{br,h}|\leq \overline{P}^{Br}_{br}, \hspace{3.0cm} \forall\;br\in \mathcal{BR}, h\in\mathcal{H}\label{eq:elec_pline_limit}\\
& 0 \leq P^{D}_{b,h} \leq \overline{P}^{D}_{b,h}, \hspace{3.2cm} \forall b\in\mathcal{B}, h\in\mathcal{H}\label{eq:elec_demand}\\
&\begin{aligned}
\sum_{i\in\mathcal{I}_b}^{}{{P}^{Gen}_{i,h}}+\sum_{v\in\mathcal{PV}_b}^{}{P^{PV}_{v,h}}+\sum_{w\in\mathcal{WT}_b}^{}{P^{WT}_{w,h}}=P^{D}_{b,h}\\+\sum_{br\in\mathcal{BR}^{fr}_b}{P^{Br}_{br,h}}-\sum_{br\in\mathcal{BR}^{to}_b}{P^{Br}_{br,h}}, \hspace{1.0cm} \forall b\in\mathcal{B}, h\in\mathcal{H}\label{eq:elec_balance}\end{aligned}
\end{alignat}\vspace{-.35cm}
\end{subequations}
\par The objective given in \eqref{eq:elec_obj} aims to minimize generation costs while serving as much demand as possible. The first term in the objective minimizes the total generation cost based on a quadratic function. The second term in the objective represents the penalty for not serving the requested demand. As shown in \eqref{eq:elec_dc}, the power flow through each branch is a function of {the} voltage angles of the two ending buses of that branch. Equations \eqref{eq:elec_p_limit}, \eqref{eq:elec_pline_limit}, and \eqref{eq:elec_demand} set bounds for {the} generation of units, flowing power through lines, and the served demand, respectively. Finally, the nodal balance equation is given in \eqref{eq:elec_balance}.
\vspace{.15cm}
\subsubsection{Natural Gas Network Look-Ahead Operation Problem}
Under the isothermal condition assumption, the one-dimensional gas pipe dynamics through horizontal pipelines are presented as PDEs in \eqref{eq:pde}.
\vspace{-.1cm}
\begin{subequations} \label{eq:pde}
\begin{alignat}{3}
&\frac{\partial }{\partial t}p(x,t)+\frac{4c^2}{\pi D^2}\frac{\partial }{\partial x}m(x,t)=0 \label{eq:pde1}\\
&\begin{aligned}
    \frac{\partial }{\partial x}p(x,t)+\frac{4}{\pi D^2}\frac{\partial }{\partial t}m(x,t)
+\frac{\partial }{\partial x}\rho(x,t)u^2(x,t)
= \\-\frac{8fc^2}{\pi^2D^5}\frac{m^2(x,t)}{p(x,t)} \label{eq:pde2}
\end{aligned}
\vspace{-0.1cm}
\end{alignat}
\end{subequations}
\par In \eqref{eq:pde}, the spatial (indexed by $x$) and temporal (indexed by $t$) characteristics of the gas pressure ($p$), the mass flow rate of the gas ($m$), gas density ($\rho$), and gas flow velocity ($u$)  inside the pipeline are related to each other by partial derivations over time and space. By assuming time-invariant gas injections, the dynamics inside gas pipes disappear, and the Weymouth equation is reached. As seen in \eqref{eq:weymouth}, {the} Weymouth equation models the average volume flow within a pipe ($Q_{mn}$) as a function of the pressures at the ending junctions of that pipe ($p_m,p_n$). Here, $k_{mn}$ is a constant based on the pipe characteristics.
\begin{equation}
    Q^2_{mn}=k_{mn}(p^2_m-p^2_n)
    \label{eq:weymouth}
\end{equation}

However, the Weymouth equation may not reliably model gas dynamics. Thus, this work is based on the transient model to ensure that ﬂuid dynamics are primarily considered.  By neglecting the term $\frac{\partial }{\partial x}\rho(x,t)u^2(x,t)$ in \eqref{eq:pde} and implementing the finite difference method \cite{liu2019optimal}, the equations for a natural gas network are presented as part of the non-convex OGF problem \eqref{eq:gas_side}. 
\begin{subequations} \label{eq:gas_side}
\begin{alignat}{3}
&\underset{v^{G},d}{\min}\sum_{t\in\mathcal{T}}\sum_{g\in\mathcal{G}}^{}{\xi_{g}^{t} v_{g,t}^{G}}+\kappa_{G}\sum_{t\in\mathcal{T}}\sum_{u\in\mathcal{U}}{(F^G_u(P^{U}_{u,t})-d_{u,t})}\label{eq:gas_side_obj}\\
& \textit{subject to:} \nonumber \\
&\underline{pr}_j^{J}\leq pr^J_{j,t} \leq \overline{pr}_j^{J},\;\;\;\;\;\;\;\;\forall j \in \mathcal{J}, t\in\mathcal{T}\;\;\;:\;\;\; \underline{\mu}^P_{j,t},\overline{\mu}^P_{j,t} \label{eq:gas_side_prj_limit}\\
&\underline{v}_g^{G} \leq v_{g,t}^{G} \leq \overline{v}_g^{G},\;\;\;\;\;\;\;\;\;\;\;\;\forall g \in \mathcal{G} , t\in\mathcal{T}\;\;\;:\;\;\; \underline{\mu}^G_{g,t},\overline{\mu}^G_{g,t}\label{eq:gas_side_gs_limit}\\
& pr^J_{j,t} = pr_{p,{1}}^{t},\;\;\;\;\;\forall j \in \mathcal{J}\setminus\mathcal{C},p\in \mathcal{P}^{fr}_j, t\in\mathcal{T}\;\;\;:\;\;\; \lambda^{fr}_{p,t}\label{eq:gas_side_pr_from}\\
& pr^J_{j,t} = pr_{p,{n}^{seg}_p}^{t},\;\;\;\;\;\;\forall j \in \mathcal{J},p\in \mathcal{P}^{to}_j, t\in\mathcal{T}\;\;\;:\;\;\; \lambda^{to}_{p,t}\label{eq:gas_side_pr_to}\\
& pr^J_{j,t} \leq pr_{p,{1}}^{t} \leq \Gamma.pr^J_{j,t} ,\forall j \in \mathcal{C},p\in \mathcal{P}^{fr}_j, t\in\mathcal{T}: \underline{\mu}^{c}_{p,t},\overline{\mu}^{c}_{p,t}\label{eq:gas_side_pr_comp}\\
&0 \leq d_{u,t}\leq F^G_u(P^{U}_{u,t}),\;\;\;\forall u\in\mathcal{U}, t\in\mathcal{T} \;\;\;:\;\;\;\underline{\mu}^U_{u,t},\overline{\mu}^U_{u,t}\label{eq:gas_side_served}\\
&\begin{aligned}\sum_{g\in\mathcal{G}_j}v_{(g,t)}^{G}+\sum_{p\in\mathcal{P}^{to}_j} m^{t}_{p,{n}^{seg}_p}-\sum_{p\in\mathcal{P}^{fr}_j} m^{t}_{p,{1}}=\sum_{l\in\mathcal{L}_j} d^G_{l,t}
,\;\;\; \\+\sum_{u\in\mathcal{U}_j}d_{u,t},\;\;\;\;\;\; \forall j \in \mathcal{J} , t\in\mathcal{T} \;\;\;:\;\;\;\lambda^L_{j,t}\label{eq:gas_side_balance}\end{aligned}\\
&\begin{aligned} \dfrac{pr^{t+1}_{p,s+1}-pr^{t}_{p,s+1}}{2\Delta t} +\dfrac{pr^{t+1}_{p,s}-pr^{t}_{p,s}}{2\Delta t} 
+\dfrac {m^{t+1}_{p,s+1}-m^{t+1}_{p,s}}{(\pi D^2/4c^2)\Delta x}\\
=0,\;\;\;\;\forall p \in \mathcal{P}, s \in \mathcal{S}_p, t\in\mathcal{T}\;\;\;:\;\;\;\eta^t_{p,s}\label{eq:gas_side_dyna1}
\end{aligned}\\
&\begin{aligned} \dfrac{pr^{t+1}_{p,s+1}-pr^{t+1}_{p,s}}{\Delta x}
+\dfrac {m^{t+1}_{p,s+1}-m^{t}_{p,s+1}}{(\pi D^2/2)\Delta t}
+\dfrac {m^{t+1}_{p,s}-m^{t}_{p,s}}{(\pi D^2/2)\Delta t}\\
+\dfrac{{m^{t}_{p,s}}^2}{(\pi^2D^5/8fc^2)pr^{t}_{p,s}}=0,\;\;\;\;\forall p \in \mathcal{P}, s \in \mathcal{S}_p, t\in\mathcal{T}\label{eq:gas_side_dyna2}\end{aligned}
\end{alignat}
\end{subequations}

According to \eqref{eq:gas_side_obj}, the objective function is to minimize the total cost of natural gas suppliers, which is composed of two elements. The first term is the production cost of natural gas, which is obtained with a linear cost function. The second term in \eqref{eq:gas_side_obj} penalizes the amount of natural gas fuel required by a unit to deliver its scheduled generation but is not served. The upper-bound and lower-bound for the output of natural gas suppliers and the pressure {at} each junction, at each time $t$, are set in \eqref{eq:gas_side_prj_limit} and \eqref{eq:gas_side_gs_limit}, respectively. The pressures at the first and last segments of each pipe are considered to be equal to the pressure {at} the junction adjacent to that segment, as shown in \eqref{eq:gas_side_pr_from} and \eqref{eq:gas_side_pr_to}, respectively. Based on  parameter $\Delta x$ and the pipe's length, every pipe in the natural gas network is sectionalized into several segments indexed by $s$. The compressor model is presented in \eqref{eq:gas_side_pr_comp}. This equation ensures that the pressure can be increased up to $\Gamma$ times in the compressor junctions, which is a parameter greater than 1. It is assumed that the natural gas demand of the gas-fired generators can be shed, according to \eqref{eq:gas_side_served}. Natural gas supply-demand balance for each junction is ensured in \eqref{eq:gas_side_balance}.  The natural gas dynamics are captured in \eqref{eq:gas_side_dyna1} and \eqref{eq:gas_side_dyna2}. The last term in \eqref{eq:gas_side_dyna2} includes a non-linear quadratic fraction that makes the problem presented in \eqref{eq:gas_side} non-linear and non-convex. Throughout this paper, the term `non-linear model/problem' refers to the original non-linear non-convex OGF problem presented above.

\section {Solution Methodology}
\begin{figure}[h!]
\vspace{-1.05cm}
    \includegraphics[width=\linewidth]{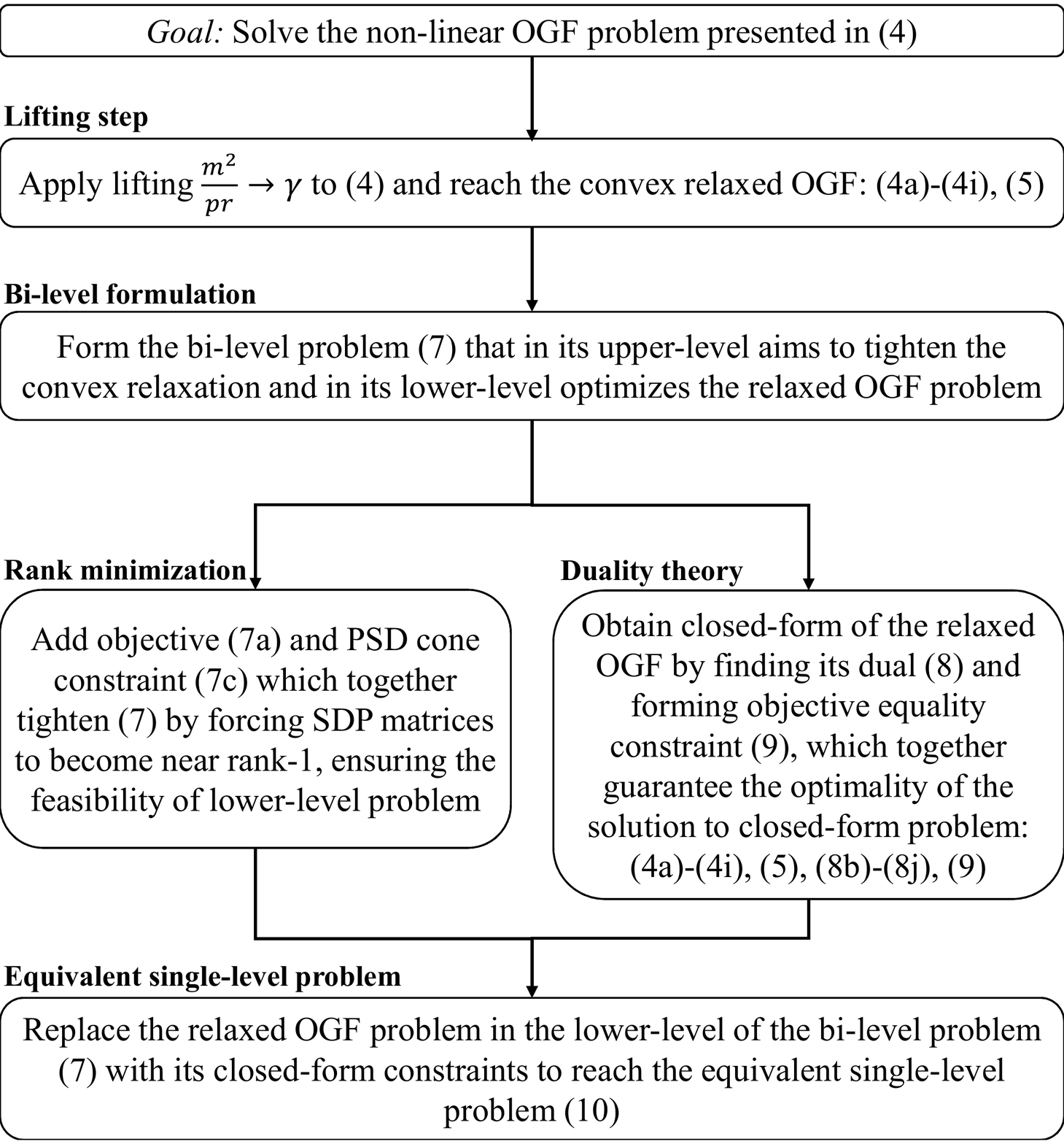}
    \vspace{-1.55cm}
    \caption{ The flowchart of the proposed rank minimization scheme }
    \label{fig:flowchart_method}
    \vspace{-0.65cm}
\end{figure}

In this section, the procedure by which we solve the non-convex OGF problem is described step by step in detail. First, the relaxation scheme to obtain a convex problem is discussed. Later, it is shown how we ensure obtaining a feasible solution to the relaxed problem by forming a bi-level optimization problem.  A rank minimization technique is introduced in the upper-level problem to guarantee the tightness of the procured optimal solution in the lower-level problem. In the end, to solve the bi-level problem, an equivalent single-level problem is presented by leveraging the dual form of the lower-level problem. An overview of the proposed methodology is displayed in Fig. \ref{fig:flowchart_method}.

\subsection{The Proposed Relaxation Scheme}
To deal with the non-convexity issue raised in Section II, we introduce the lifting variable ${(m^{t}_{p,s})^2}/{pr^{t}_{p,s}}\overset{lift}{\longrightarrow}\gamma^t_{p,s}$ to relax the non-convex term within the natural gas dynamics equation. As a result, \eqref{eq:gas_side_dyna2} is converted to the relaxed equation introduced by \eqref{eq:gas_side_gamma}.
\begin{equation}\label{eq:gas_side_gamma}
\begin{aligned}
    \dfrac{pr^{t+1}_{p,s+1}-pr^{t+1}_{p,s}}{\Delta x}
    +\dfrac {m^{t+1}_{p,s+1}-m^{t}_{p,s+1}}{(\pi D^2/2)\Delta t}
    +\dfrac {m^{t+1}_{p,s}-m^{t}_{p,s}}{(\pi D^2/2)\Delta t}\\
    +\dfrac{\gamma^t_{p,s}}{(\pi^2 D^5/8fc^2)}=0;\;\;\;\;\forall p \in \mathcal{P}, s \in \mathcal{S}_p, t\in\mathcal{T}\;\;:\;\;\zeta^t_{p,s}
\end{aligned}
\end{equation}

The lifting step relaxes the original dynamic OGF problem \eqref{eq:gas_side} as it deals with the non-linear term. However, the solution procured by the relaxed problem might not be {necessarily} feasible. The $\gamma^{t}_{p,s}$ acquired from the primal relaxation problem presented in \eqref{eq:gas_side_prj_limit}-\eqref{eq:gas_side_dyna1} and \eqref{eq:gas_side_gamma} cannot guarantee that $m^{t}_{p,s}$ and $pr^{t}_{p,s}$ can be uniquely obtained. The bi-level optimization problem is presented here to resolve the issue. The rank is minimized in the upper-level problem and solved to optimality in the lower-level problem so that the procured solution is feasible and optimal.

\vspace{-.35cm}\subsection{Rank Minimization via Bi-level Formulation }
In the standard rank minimization approach, the relaxed problem is solved by addition of a penalty term to the objective function \cite{sojoudi2013low,madani2014convex,fazelnia2016convex,liu2017rank}. Whereas in the presented work, a bi-level problem is presented which minimizes rank in the objective of its upper-level problem and considers the relaxed problem in its lower level. Doing so changes the problem structure and improves the tightness of the solution significantly. The upper-level problem of the formulated bi-level problem aims to enforce the tightness of the presented relaxation scheme. The optimality of the procured solution is also achieved in the lower-level problem.  To achieve the desired tightness, the matrix formed in \eqref{eq:sdp_matrix} should be {rank-1} for all segments of all pipes in the natural gas infrastructure at all times.
\begin{equation}\label{eq:sdp_matrix}
\begin{aligned}
    \begin{bmatrix}
    pr^{t}_{p,s} & m^{t}_{p,s}\\ 
    m^{t}_{p,s} & \gamma^t_{p,s}
    \end{bmatrix} , \;\;\;\;\forall p \in \mathcal{P}, s \in \mathcal{S}_p, t\in\mathcal{T}
\end{aligned}
\end{equation}

To obtain a tight near-rank-1 solution for the relaxed natural gas operation problem, an objective function that minimizes the lifting variable $\gamma^t_{p,s}$ is assigned to the upper-level problem, and a Positive Semi-Definite (PSD) constraint is added.
The PSD constraint connects the principle terms $pr^{t}_{p,s}$ and $m^{t}_{p,s}$ with the lifting term $\gamma^t_{p,s}$. If the cone matrix is rank-1, the lifting term accurately represents the relationship between the {principal} elements. The obtained bi-level problem is presented in \eqref{eq:bi_level}. \vspace{-.35cm}
\begin{subequations} \label{eq:bi_level}
\begin{alignat}{3}
&\begin{aligned}\underset{\gamma}{\min}\sum_{t\in\mathcal{T}}\sum_{p\in\mathcal{P}}\sum_{s\in\mathcal{S}_p}\gamma^t_{p,s}\label{eq:bi_level_obj}\end{aligned}\\
& \nonumber \textit{subject to: }\\
&  \gamma \in \text{argmin } \{\eqref{eq:gas_side_obj} \textit{ subject to: \eqref{eq:gas_side_prj_limit}-\eqref{eq:gas_side_dyna1}, \eqref{eq:gas_side_gamma}}\} \label{eq:bi_level_lower} \\
&\begin{bmatrix}
    pr^{t}_{p,s} & m^{t}_{p,s}\\ 
    m^{t}_{p,s} & \gamma^t_{p,s}
    \end{bmatrix} \succeq 0, \;\;\;\;\forall p \in \mathcal{P}, s \in \mathcal{S}_p, t\in\mathcal{T}\label{eq:bi_level_psd}
\end{alignat}
\end{subequations}

Here, the objective function aims to minimize the lifting variable $\gamma$. The minimization objective and the PSD constraint force the SDP matrices to become rank-1 to ensure the tightness of the obtained solution. The rank of the rendered SDP matrices is near one, i.e. the largest eigenvalue of each matrix dominates the rest of them.  It is noteworthy to mention that the SDP cone in \eqref{eq:bi_level_psd} is of size $2\times 2$, which is mathematically equivalent to {the} SOC constraint. Using SOC relaxation instead of SDP could result in marginally faster solution times at the price of slightly less tight relaxation. To solve the bi-level problem \eqref{eq:bi_level}, the lower-level problem should be presented as a set of constraints (i.e., its closed-form) by obtaining the dual of the relaxed problem and employing the primal-dual constraint equalizing the objectives of {the} primal and dual problems. The fundamental differences of a bi-level optimization problem and a multi-objective problem are discussed in  \cite{glackin2009solving}.

\par Our novelty compared with the standard rank minimization approaches \cite{sojoudi2013low,madani2014convex,fazelnia2016convex,liu2017rank,manshadi2019convex} is the employment of the bi-level presentation. In standard rank minimization methods, the objective function \eqref{eq:gas_side_obj} is perturbed by {the} addition of the penalty term \eqref{eq:bi_level_obj}. This method leads to a multi-objective single-level problem and may not result in a tight relaxation. With changes in the weights of each objective, different solutions are obtained. To solve this issue, a bi-level problem formulation is introduced that emphasizes on the relaxation tightness. The rank minimization objective is placed in the upper-level problem to ensure the tightness of the solution. Hence, the upper-level OGF problem returns feasible and physically meaningful solutions {to} the original non-convex problem. Also, the lower-level objective aims to minimize the cost of the OGF problem, meaning the procured solution is optimal.

\subsection{The Closed-Form Representation of the Relaxed OGF Problem}
By relaxing the non-linear term in the original OGF problem, a convex formulation is reached. This allows employing the dual form of the relaxed problem and presenting the lower-level problem as a set of constraints. To obtain the closed-form of the relaxed OGF problem, the constraints of the primal and dual problems along with {the} primal-dual objective pairing constraint are employed. In front of each constraint in the primal problem, the corresponding dual variables are symbolized, and vice versa. The dual form of the relaxed primal problem presented by \eqref{eq:gas_side_obj}-\eqref{eq:gas_side_dyna1} and \eqref{eq:gas_side_gamma} is formulated in \eqref{eq:dual}.
\begin{subequations} \label{eq:dual}
\begin{alignat}{3}
&\begin{aligned}\underset{\lambda,\underline{\mu},\overline{\mu}}{\max}\sum_{t\in\mathcal{T}}\sum_{j\in\mathcal{J}}\lambda^L_{j,t}\sum_{l\in\mathcal{L}_j} d^G_{l,t}+ \sum_{t\in\mathcal{T}}\sum_{g\in\mathcal{G}}(\underline{v}_g^{G}\underline{\mu}^G_{g,t}-\overline{v}_g^{G}\overline{\mu}^G_{g,t})
\\-\sum_{t\in\mathcal{T}}\sum_{u\in\mathcal{U}}{F^G_u(P^{U}_{u,t})\overline{\mu}^U_{u,t}}+ \sum_{t\in\mathcal{T}}\sum_{j\in\mathcal{J}}(\underline{pr}_j^{J}\underline{\mu}^P_{j,t}-\overline{pr}_j^{J}\overline{\mu}^P_{j,t}) \label{eq:dual_obj}\end{aligned}\\
& \textit{subject to:} \nonumber\\
&\underline{\mu}^U_{u,t}-\overline{\mu}^U_{u,t}-\lambda^L_{{j}_u,t}\leq -\kappa_G,\label{eq:dual_served}\hspace{.5cm}\forall u \in \mathcal{U}, t\in\mathcal{T}\;:\;d_{u,t}\\
&\begin{aligned}\underline{\mu}^P_{j,t}-\overline{\mu}^P_{j,t}+\lambda^{to}_{{p}^{to}_j,t}+\lambda^{fr}_{{p}^{fr}_j,t}\leq 0,\hspace{3cm}\\\label{eq:dual_prj}\;\forall j \in \mathcal{J}\setminus\mathcal{C}, t\in\mathcal{T}\;:\;pr^J_{j,t}\end{aligned}\\
&\begin{aligned}\underline{\mu}^P_{j,t}-\overline{\mu}^P_{j,t}+\lambda^{to}_{{p}^{to}_j,t}+\underline{\mu}^{c}_{{p}^{fr}_j,t}-\Gamma.\overline{\mu}^{c}_{{p}^{fr}_j,t}\leq 0,\hspace{1.5cm}\\\label{eq:dual_prj_comp}\;\;\forall j \in \mathcal{C}, t\in\mathcal{T}\;:\;pr^J_{j,t}\end{aligned}\\
&\lambda^L_{j,t}\geq 0,\label{eq:dual_fin}\;\;\;\;\;\;\;\;\;\;\;\;\;\;\;\;\;\;\;\;\;\;\;\;\;\;\;\forall j \in \mathcal{J}^{fr}_p, t\in\mathcal{T}\;\;\;:\;\;\;f^i_{p,t}\\
&\lambda^L_{j,t}\leq 0,\label{eq:dual_fout}\;\;\;\;\;\;\;\;\;\;\;\;\;\;\;\;\;\;\;\;\;\;\;\;\;\;\forall j \in \mathcal{J}^{to}_p, t\in\mathcal{T}\;\;\;:\;\;\;f^o_{p,t}\\
&\underline{\mu}^G_{g,t}-\overline{\mu}^G_{g,t}+\lambda^L_{j,t} =\xi_{g}^{t},\label{eq:dual_vgs}\forall g \in \mathcal{G}, j \in \mathcal{J}_g, t\in\mathcal{T}\;:\; v^G_{j,t}\\
&\begin{aligned} \dfrac{\eta^{t-1}_{p,s-1}-\eta^{t}_{p,s-1}}{2\Delta t} +\dfrac{\eta^{t-1}_{p,s}-\eta^{t}_{p,s}}{2\Delta t} + \dfrac{\zeta^{t-1}_{p,s-1}-\zeta^{t-1}_{p,s}}{\Delta x}\leq 0,\\ \;\;\;\;\forall p \in \mathcal{P}, s \in \mathcal{S}_p, t\in\mathcal{T}\;\;\;:\;\;\;pr^t_{p,s}\label{eq:dual_pressure}
\end{aligned}\\
&\begin{aligned} 
\dfrac {\eta^{t-1}_{p,s-1}-\eta^{t-1}_{p,s}}{(\pi D^2/4c^2)\Delta x}
+\dfrac {\zeta^{t-1}_{p,s-1}-\zeta^{t}_{p,s-1}}{(\pi D^2/2)\Delta t}
+\dfrac {\zeta^{t-1}_{p,s}-\zeta^{t}_{p,s}}{(\pi D^2/2)\Delta t}
\leq 0,\\ \;\;\;\;\forall p \in \mathcal{P}, s \in \mathcal{S}_p, t\in\mathcal{T}\;\;\;:\;\;\;m^t_{p,s}\label{eq:dual_mass}\end{aligned}\\
&\zeta^t_{p,s}\leq 0,\label{eq:dual_gamma}\;\;\;\;\;\;\;\;\;\;\;\;\;\;\;\;\;\;\forall p \in \mathcal{P}, s \in \mathcal{S}_p, t\in\mathcal{T}\;\;\;:\;\;\;\gamma^t_{p,s}
\end{alignat}
\end{subequations}

In \eqref{eq:dual_served}-\eqref{eq:dual_gamma}, the corresponding variables for which the dual is obtained are presented in front of each constraint.
The primal-dual equality constraint is presented in \eqref{eq:equal_primal_dual_obj} to achieve the closed-form of the lower-level problem, which is added to the set of constraints presenting the closed-form of the lower-level problem \eqref{eq:gas_side_prj_limit}-\eqref{eq:gas_side_dyna1}, \eqref{eq:gas_side_gamma}, and  \eqref{eq:dual_served}-\eqref{eq:dual_gamma}. Thus, the lower-level problem which ensures the optimality of the OGF problem is presented in the closed-form by \eqref{eq:gas_side_prj_limit}-\eqref{eq:gas_side_dyna1}, \eqref{eq:gas_side_gamma},  \eqref{eq:dual_served}-\eqref{eq:dual_gamma}, \eqref{eq:equal_primal_dual_obj}. 
\begin{equation} \label{eq:equal_primal_dual_obj}
\begin{aligned}\sum_{t\in\mathcal{T}}\sum_{j\in\mathcal{J}}\lambda^L_{j,t}\sum_{l\in\mathcal{L}_j} d^G_{l,t}+ \sum_{t\in\mathcal{T}}\sum_{g\in\mathcal{G}}(\underline{v}_g^{G}\underline{\mu}^G_{g,t}-\overline{v}_g^{G}\overline{\mu}^G_{g,t})
\\ + \sum_{t\in\mathcal{T}}\sum_{j\in\mathcal{J}}(\underline{pr}_j^{J}\underline{\mu}^P_{j,t}-\overline{pr}_j^{J}\overline{\mu}^P_{j,t})-\sum_{t\in\mathcal{T}}\sum_{u\in\mathcal{U}}{F^G_u(P^{U}_{u,t})}=\\\sum_{t\in\mathcal{T}}\sum_{g\in\mathcal{G}}^{}{\xi_{g}^{t} v_{g,t}^{G}}+\kappa_G\sum_{t\in\mathcal{T}}\sum_{u\in\mathcal{U}}{(F^G_u(P^{U}_{u,t})-d_{u,t})}\end{aligned}
\end{equation}

\subsection{The Equivalent Single-level Rank Minimization Problem for the Relaxed OGF Problem}

Finally, by transforming the lower-level problem into a set of constraints, an equivalent single-level problem is obtained with {the} general form displayed in \eqref{eq:eq_1level}.
\begin{subequations} \label{eq:eq_1level}
\begin{alignat}{3}
& \underset{\gamma}{\min}\sum_{t\in\mathcal{T}}\sum_{p\in\mathcal{P}}\sum_{s\in\mathcal{S}_p}\gamma^t_{p,s} \\
& \nonumber \textit{subject to:}\\
& \eqref{eq:gas_side_prj_limit}-\eqref{eq:gas_side_dyna1}, \eqref{eq:gas_side_gamma}, \eqref{eq:bi_level_psd}, \eqref{eq:dual_served}-\eqref{eq:dual_gamma}, \eqref{eq:equal_primal_dual_obj},
\end{alignat}
\end{subequations}

The presented problem \eqref{eq:eq_1level} is a single-level optimization with sparse SDP cones related to each segment of each pipeline at each time. As the computation order of the SDP problem is associated with the size of these 2$\times$2 matrices and the size of each matrix will not change with {the} scaling of the network, the proposed approach will be able to scale for problems associated with various network sizes efficiently.

\section {Results and Discussions}
This section comprises two cases. In the first case, the operation problem in a sample interdependent network including a 6-bus electricity network and a 6-junction natural gas network is presented. In this case, two critical discussions are presented. First, it is discussed why the Weymouth equation is not suitable for the short-term operation problem, as it leverages approximations that are based on conditions that do not hold in shorter periods.
Second, the merit of the proposed methodology is demonstrated according to the results obtained by applying the proposed model. It is shown that with the proposed relaxation model, a tight optimal solution is procured, which requires substantially lower solver time than the original dynamics model. 
In the second case, the results of applying the proposed model to a  larger  network are displayed. This network consists of the modified IEEE 118 bus network combined with a 10 junction natural gas network. For the original non-convex OGF problem and the proposed model, the time step length is 5 minutes, i.e., $\Delta t=300~s$, and the length of pipe segments is  5 kilometers, i.e., $\Delta x=5000~m$. All simulations are performed on a PC with {an} Intel Core i7 3.60GHz CPU using Julia programming language \cite{bezanson2017julia} in the JuMP \cite{DunningHuchetteLubin2017} environment. The non-convex model is solved by IPOPT \cite{wachter2006implementation}, while the MOSEK\cite{mosek} solver is used to solve the proposed formulation. The required data for replication of the results in case studies {is} available at \cite{data_github}.

\begin{figure}[h!]
    \centering
    \vspace{-0.45cm}
    \includegraphics[width=\linewidth]{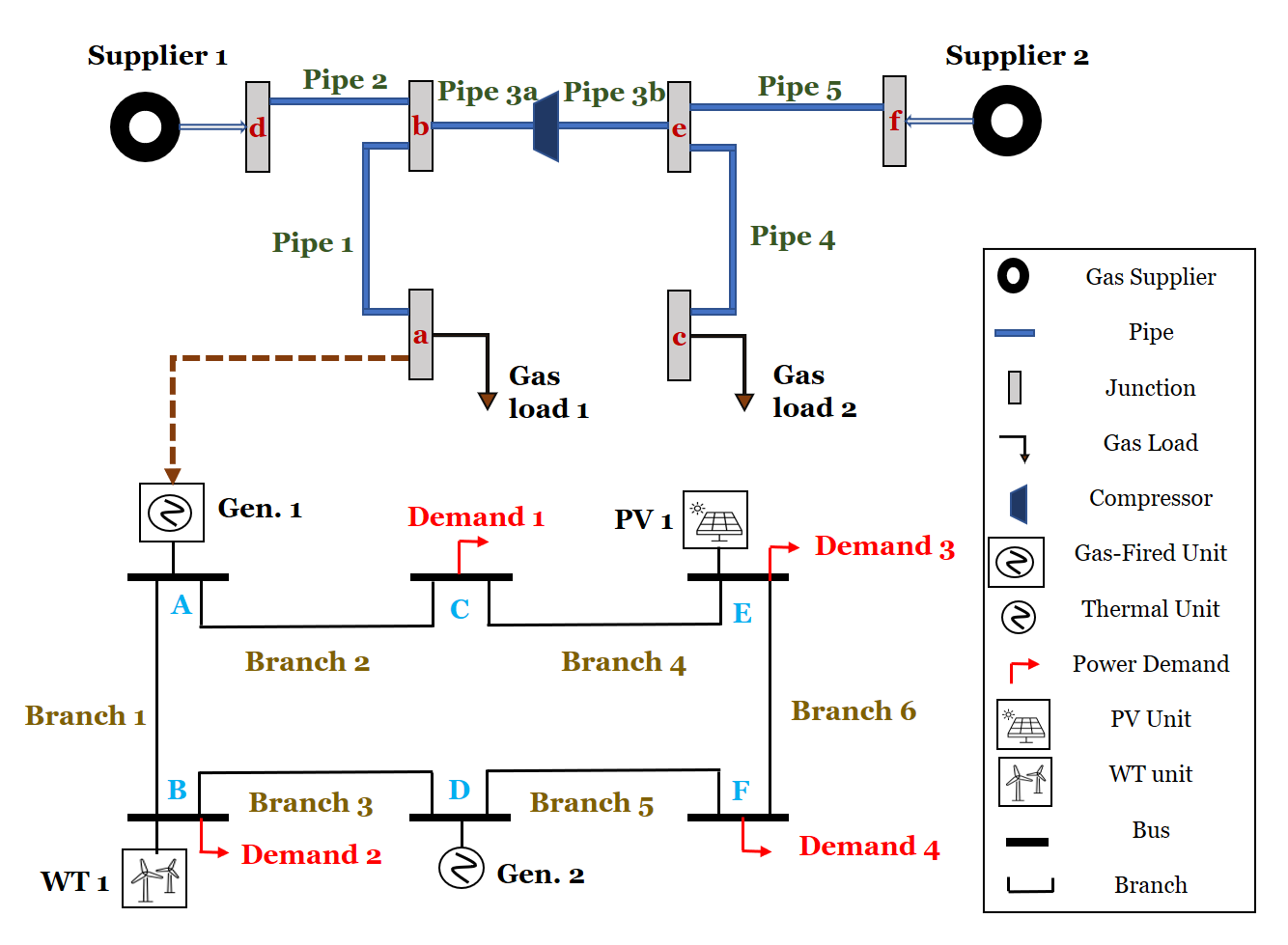}
    \vspace{-0.75cm}
    \caption{6-Junction natural gas network coupled with 6-bus power network}
    \label{fig:coupled_6}
    \vspace{-0.15cm}
\end{figure}

\subsection{A Sample Interdependent Electricity and Natural Gas Network}
\par First, the incapability of the Weymouth model in short-term operation scenarios is illustrated  and compared against the solution of the non-convex model. Then, the accuracy and exactness of the proposed relaxation method are displayed. The configuration of the test 6-junction natural gas network coupled with a 6-bus power network is displayed in Fig. \ref{fig:coupled_6}. The natural gas network serves two types of demands. First, two time-varying heat demands with peak {values} of 1000 $(kcf/hr)$ and 500 $(kcf/hr)$ are to be fed at junctions $a$ and $c$, respectively. These demand points have a high priority and must be served at all times. The second type of demand served by the natural gas network is the natural gas required for the operation of gas-fired units within the 6-bus power network. This demand is placed on junction `a' and is of lower priority than the heat demand so that it can be shed. The specifications of the natural gas network and suppliers are displayed in Table \ref{table:6_ng_network}. A compressor with a maximum compression ratio of 1.1 is also placed on pipe 3. Pipe 3 is represented {by two} sections to distinguish between pressures at the two ends of the compressor.
\begin{table}[h!]
\vspace{-0.35cm}
	\footnotesize \centering
	\caption {6-Junction natural gas network specifications} 
	\vspace{-0.2cm}
\begin{tabular}{cccc} \hline\hline
\textbf{Pipelines}\\ \hline
\multicolumn{1}{c}{Pipe}  & \multicolumn{1}{c}{From} & \multicolumn{1}{c}{To} & Length (km) \\  \hline
1 & {\textit{b}} & {\textit{a}} & 100 \\
2 & {\textit{d}} & {\textit{b}} & 80 \\
3 & {\textit{e}} & {\textit{b}} & 120 \\
4 & {\textit{e}} & {\textit{c}} & 100 \\
5 & {\textit{f}} & {\textit{e}} & 80 \\
\hline\hline
\textbf{Suppliers}                \\ \hline
{Number} & $\underline{v}^G$ ($kcf/hr$) & $\overline{v}^G$ ($kcf/hr$) & Cost ($\$/kcf$) \\ \hline
1 & 125 & 875 & 1.2\\
2 & 200 & 1100 & 1\\\hline\hline
\label{table:6_ng_network}
\vspace{-0.45cm}
\end{tabular}
\end{table}
\subsubsection{\textbf{Limitation of the Weymouth equation in the natural gas short-term operation problem}}
\begin{figure}[h!]
    \vspace{-0.15cm}
    \centering
    \includegraphics[width=.9\linewidth]{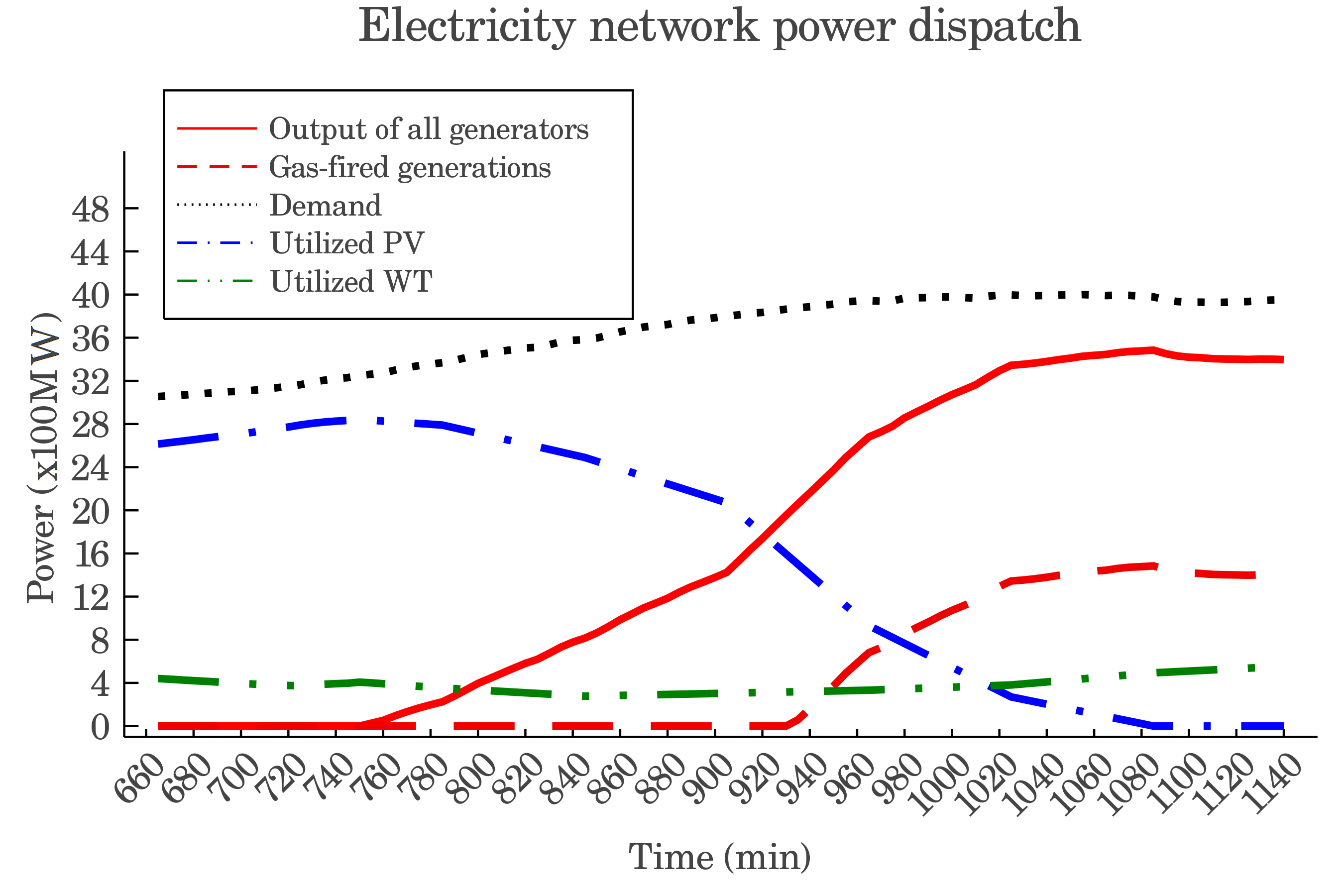}
    \vspace{-0.35cm}
    \caption{Electric power dispatch of units in the 6-bus power network}
    \label{fig:case1_dispatch}
    \vspace{-0.25cm}
\end{figure}

A short-term scenario with a varying gas load is generated to exhibit the shortcomings of the Weymouth equation in modeling natural gas network dynamics. Then, the OGF equations are formed and solved for both the non-linear OGF problem presented in \eqref{eq:gas_side} and the Weymouth model given in \eqref{eq:weymouth} as presented in \cite{manshadi2018tight}. 
Fig. \ref{fig:case1_dispatch} illustrates the dispatch of the different units in the power network from 11 A.M. to 7 P.M. (corresponding to minutes 660 through 1140 of the day). It is supposed that this network enjoys a high penetration level of renewable generation, including solar photovoltaic (PV) and wind turbine (WT) {units}. 

\par During the first hour of operation, the renewable units can provide the required demand. However, as time passes, the electricity demand gradually increases, and {the} PV generation drops. Consequently, the thermal unit output is gradually increased, to the point where it cannot meet the growing demand and the gas-fired unit comes online. A similar pattern happens almost every day inside power networks {that} contain both gas-fired and renewable generation units. The outcome is always the same: a surge in natural gas demand over the span of {a} few hours (refer to Fig. \ref{fig:ng_production}). For systems with high renewable penetration levels, the change in demand is more severe and stochastic, depending on the weather conditions.\\

\par
\begin{figure}[h!]
\vspace{-0.35cm}
    \centering
    \includegraphics[width=.9\linewidth]{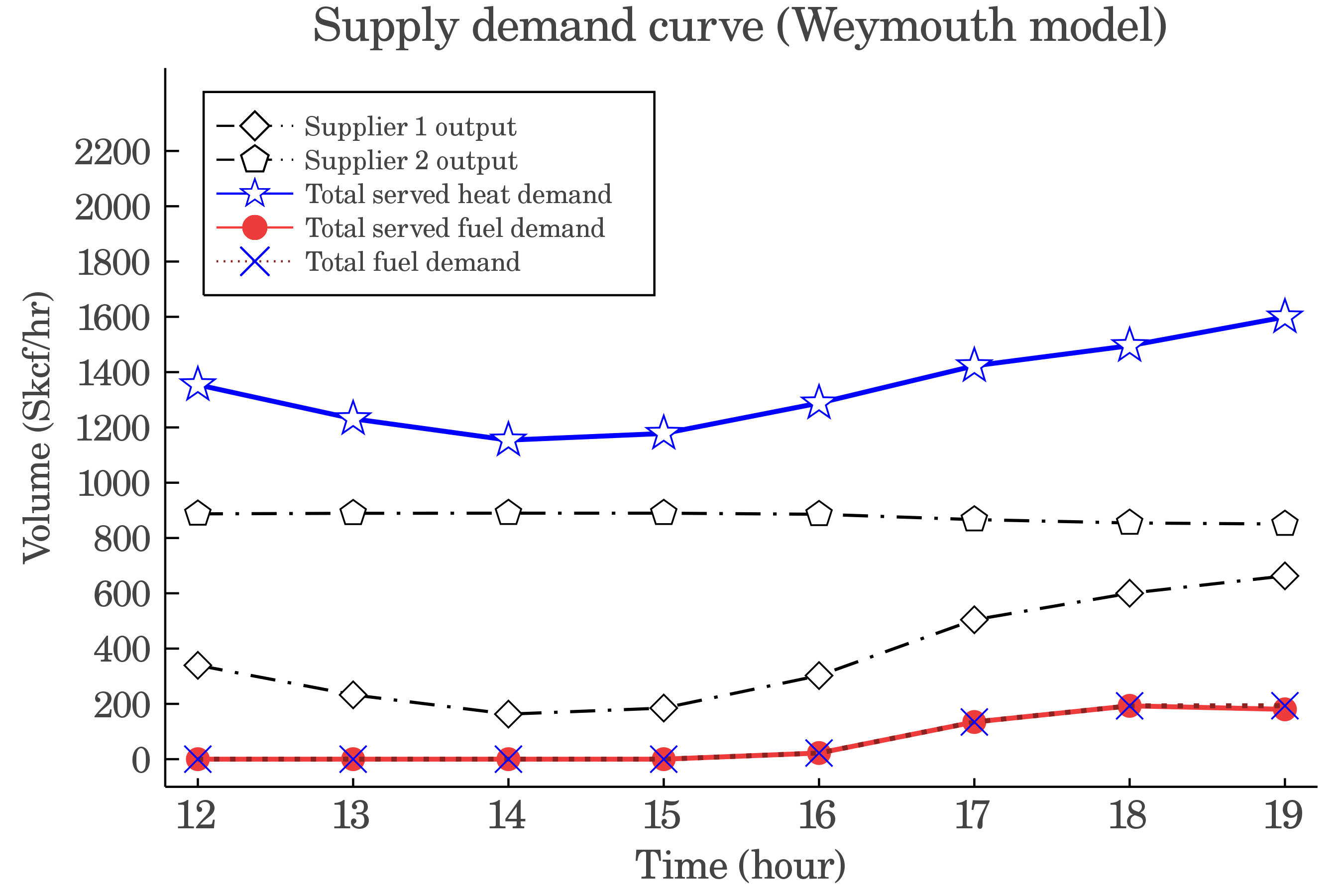}
    \vspace{-0.35cm}
    \caption{Supply demand curve of 6-junction natural gas network with Weymouth model}
    \label{fig:case1_weymouth}
    \vspace{0.35cm}
    \includegraphics[width=.9\linewidth]{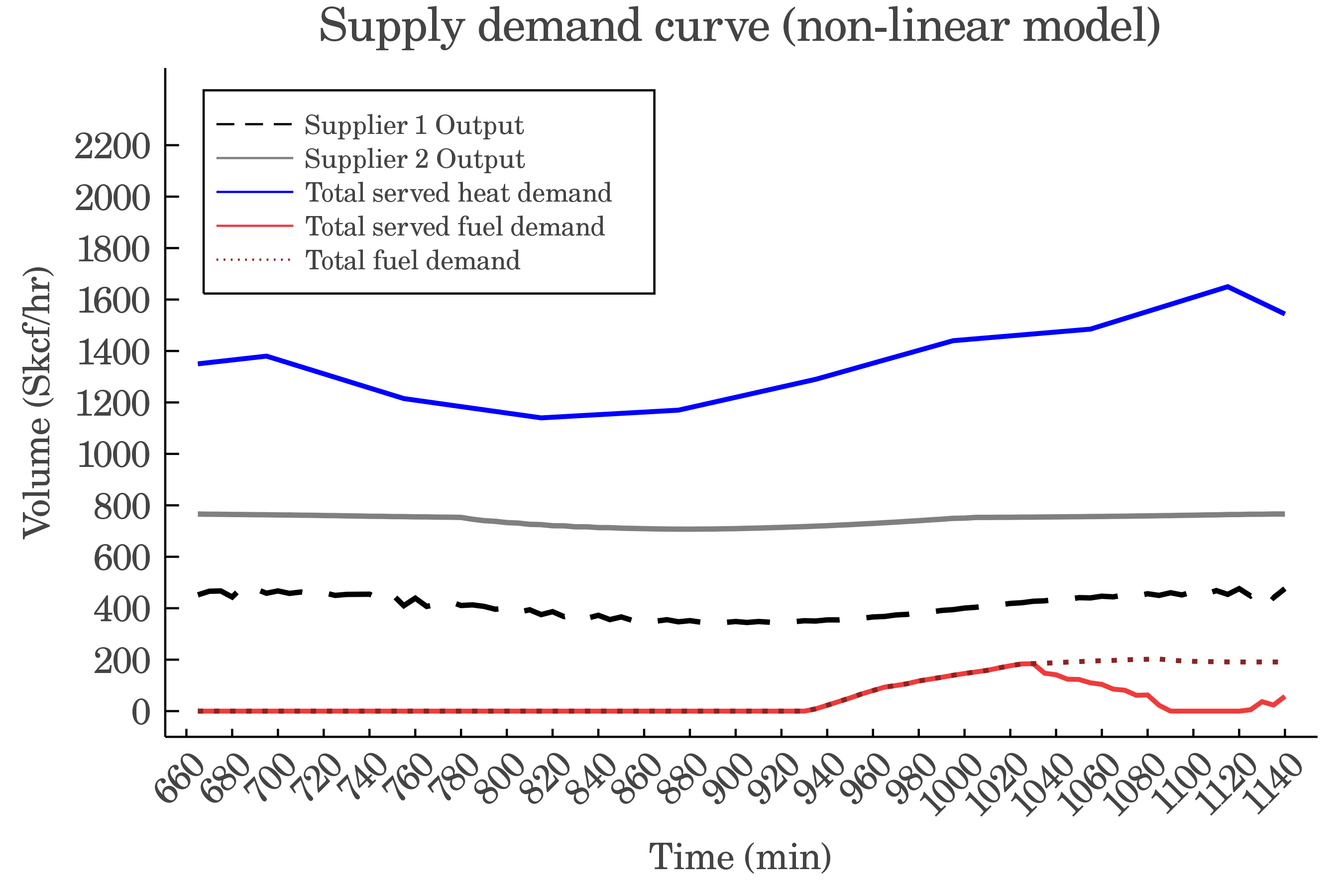}
    \vspace{-0.35cm}
    \caption{Supply demand curve of 6-junction natural gas network with non-convex model}
    \label{fig:case1_nonlinear}
    \vspace{-0.35cm}
\end{figure}

 The major drawback of employing the Weymouth model is its inability to incorporate the variations in demand and properly model their impact on the natural gas network dynamics. In addition, the Weymouth model presents an \emph{approximation} of the original natural gas dynamics model. In some cases, the combined effect of these two issues can lead to inaccurate results, which may bring about power system outages and demand curtailment. Figs. \ref{fig:case1_weymouth} and \ref{fig:case1_nonlinear} illustrate the results of the simulations performed with {the} Weymouth and original dynamics models, respectively. Comparing Fig. \ref{fig:case1_weymouth} with Fig. \ref{fig:case1_nonlinear}, the first point that stands out is the time axis. The non-convex model captures the state every 300 seconds, while the Weymouth formulation is incapable of modeling the temporal behavior of natural gas.
One assumption of the Weymouth equation is that $\Delta t \rightarrow \infty$, and here for this assumption to hold, 1-hour intervals are considered. Subsequently, only a few snapshots of the system can be chosen when implementing the Weymouth model. The non-convex model enjoys 12 intervals per hour, while the Weymouth model reports only hourly data. One consequence of this issue is missing critical information, such as the value of demand, since the Weymouth model averages over the hourly data points.
In this case, according to Fig. \ref{fig:case1_nonlinear}, the momentary demand value at minute 695 of the day as seen by the non-convex model is equal to 1380 $(Skcf/hr)$. It is observed from Fig. \ref{fig:case1_weymouth} that this value is missed with the Weymouth model, and it only observes one data point for the $12^{th}$ hour, i.e., 1354 $(Skcf/hr)$.\\

\par Utilizing the Weymouth model may {lead to} misleading solution. According to Fig. \ref{fig:case1_weymouth}, this system can fully provide the required gas fuel for the 8-hour operation period. The non-convex model, on the other hand, reports differently. According to Fig. \ref{fig:case1_nonlinear}, the system cannot meet the gas-fired unit's demand from minute 1055 to 1140 of the day. This failure is not negligible and could bring about serious fuel security issues. During these 85 minutes, more than 80\% of the desired gas for the gas-fired unit is not delivered, which will lead to inevitable outages in the electricity network.\\

\par One limitation of natural gas supply resources is their inability to respond to sudden changes in gas demand, which is a fundamental difference between the natural gas network and  the electricity network. Unlike electricity that transmits almost instantly, natural gas velocity inside the pipeline is around 40 km/h. Often, it takes a few hours for the gas supplied at the source junction to reach the demand junction. Another limitation of the Weymouth model is that it cannot properly model this characteristic of natural gas suppliers. Since the Weymouth equation is time-invariant, it is observed in Fig. \ref{fig:case1_weymouth} that the overall supplier output is constantly changing to equalize the system demand. However, the non-convex model indicates that the output of natural gas suppliers  does not vary with the change in demand. The output of supplier 1 in Fig. \ref{fig:case1_weymouth} during the hours 15 to 17 increases from 184.9 (Skcf/hr) to 504.2 (Skcf/hr). The output of the same supplier for the corresponding period according to the results of the non-convex model in Fig. \ref{fig:case1_nonlinear} increases only from 344.6 (Skcf/hr) to 418.6 (Skcf/hr). The variations in natural gas demand are addressed {by} the adjustment of the pressure inside pipelines. It is concluded that the Weymouth equation neglects the capability of natural gas pressure in dealing with variations. This observation highlights the merit of employing the proposed short-term operation model to prepare the network to hedge against uncertain demand. In a nutshell, the takeaways from the presented discussion are as follows:
\begin{itemize}[leftmargin=*]
    \item The Weymouth equation is an approximated model based on assumptions that do not hold in the short-term horizon.
    \item The Weymouth model is incapable of capturing and representing dynamic temporal  relations inside the natural gas network. For instance, the Weymouth equation equalizes natural gas demand and supply momentarily, whereas in reality, it could take a long {time} for natural gas to travel from the source junction to the demand junction.
    \item The Weymouth model misses out on all of the intra-hour information. The output power of gas-fired units {varies} distinctly due to their high ramping rates, which leads to excessive alterations in natural gas demand. Thus, applying an exact model with dynamics is crucial to capture these changes in a few minutes.
    \item The results obtained by the Weymouth model are not trustworthy in terms of natural gas fuel availability. 
    A case is designed and presented to illustrate this issue particularly. It is observed that the results of the OGF problem when applying the Weymouth model differ from those obtained by utilizing the original non-convex model with natural gas dynamics, which could result in costly consequences in the operation of a power system with several gas-fired units.
\end{itemize}

\par Utilizing a steady-state OGF model is essentially similar to neglecting the natural gas speed of delivery. The presented discussion exemplifies how overlooking the transients in the natural gas network could lead to inaccuracies that affect power system operation.

\vspace{0.3cm}
\subsubsection{\textbf{Validation of the proposed rank minimization approach}}
This part illustrates that the proposed relaxation model can incorporate dynamics of the gas network correctly by extracting a tight, optimal, and feasible solution. Unlike the Weymouth equation, this model is time-variant, which is crucial to problems in short-term time scales. Furthermore, the proposed scheme does not require long solver times, which is the major issue with non-convex models. The results of the proposed relaxation scheme are presented and compared with the original non-convex model \eqref{eq:gas_side}, the simple sparse SDP method, and the standard rank minimization relaxation method.  The simple sparse SDP method solves the lifted form of the primal side equations \eqref{eq:gas_side}, i.e., \eqref{eq:gas_side_gamma} replaces \eqref{eq:gas_side_dyna2}, with the addition of the PSD constraint \eqref{eq:bi_level_psd}.  The standard rank minimization relaxation is similar to the simple SDP relaxation but adds the penalty term \eqref{eq:bi_level_obj} to its objective function. The performance comparison of these methods is presented in Table \ref{table:validation}, where the non-convex column refers to the results of solving  problem \eqref{eq:gas_side} with an interior point solver.
\setlength\tabcolsep{2.5pt} 
\begin{table}[h!]
	\vspace{-0.3cm}
	\small \centering
	\caption {Performance and results comparison of four methods}
\resizebox{\linewidth}{!}{\begin{tabular}{ccccc} \hline\hline
\textbf{Model} & {Simple SDP} & \makecell{{Standard} \\ {Rank Minimization}}  & {Proposed} & {Non-Convex} \\  \hline
Solver time ($s$) & 0.41 &  1.04  & {1.20} & {56.8} \\ \hline
Objective ($\$$) & 19,772 &  {21,288} &  {21,343}  & {21,658} \\ \hline
Tightness ratio & {2.16} &  {6.14}  & {12.43} & - \\ \hline\hline
\label{table:validation}
\vspace{-.35cm}
\end{tabular}}
\end{table}

\par It is noticed that the proposed approach substantially decreases the solution time compared to the one required by {the} non-convex model.
The presented solution method renders a high-quality tight solution while {not} significantly increasing the computation burden of the relaxation method compared to the simple sparse SDP and standard rank minimization relaxation methods.  
It is observed that the solver time for the proposed method is only 1.2 seconds, whereas the solver time when using the exact dynamics model \eqref{eq:gas_side} is 56.8 seconds.
The solver time of the proposed model for the 24-hour simulation horizon is 13.45 seconds.  In Table \ref{table:validation}, the tightness ratios of eigenvalues for the matrix in \eqref{eq:sdp_matrix} are presented. In \cite{soofi2020socp}, the logarithmic ratio of eigenvalues of the matrix \eqref{eq:sdp_matrix}  is calculated to measure relaxation tightness. 
The tightness ratio measures the logarithmic ratio of the largest and second-largest eigenvalues of the SDP matrix. If this ratio becomes high, meaning that the second eigenvalue is dominated by the first one, the matrix can be considered rank-1. The tightness ratio is reported for all pipes, at each segment and time. The reported tightness ratio in Table \ref{table:validation} is the average for all spatio-temporal instances over the natural gas network. A tightness ratio of 12.43 means that the error for recovering the lifted variables is in {the} order of $\sim10^{-12}$.

\par Comparing the results of the simulations obtained by applying several models is almost impossible as we cannot come up with a global indicator to capture the quality of objective value, tightness, and solver time as these are different units. However, {given the low tightness ratios of other relaxation methods, {the} lower solution time and objective values of other relaxation approaches do not indicate their superiority.} The procured objective value of the natural gas operation problem presented in Table \ref{table:validation} reveals that  all three relaxation methods provide a lower bound for the objective value. However, { the critical point is that \emph{the solution procured by the proposed method is feasible for the original non-convex problem,} while this is not the case for the other relaxation schemes.} That is why the objective values of the implemented relaxation methods are not easily comparable.
{The solution rendered by the simple SDP or the standard rank minimization methods might not be physically meaningful given their relatively low tightness ratios}, i.e., lower than {the} minimum feasible solution. The one rendered by the nonlinear model is a local minimum solution, i.e., higher than the best available minimum solution. It is noteworthy to mention if SOC cones {are} used instead of SDP cones, solutions are obtained slightly faster but with a smaller tightness ratio (i.e. less reliable solutions).

 \begin{figure}[b!]
    \vspace{-0.45cm}
    \includegraphics[width=\linewidth]{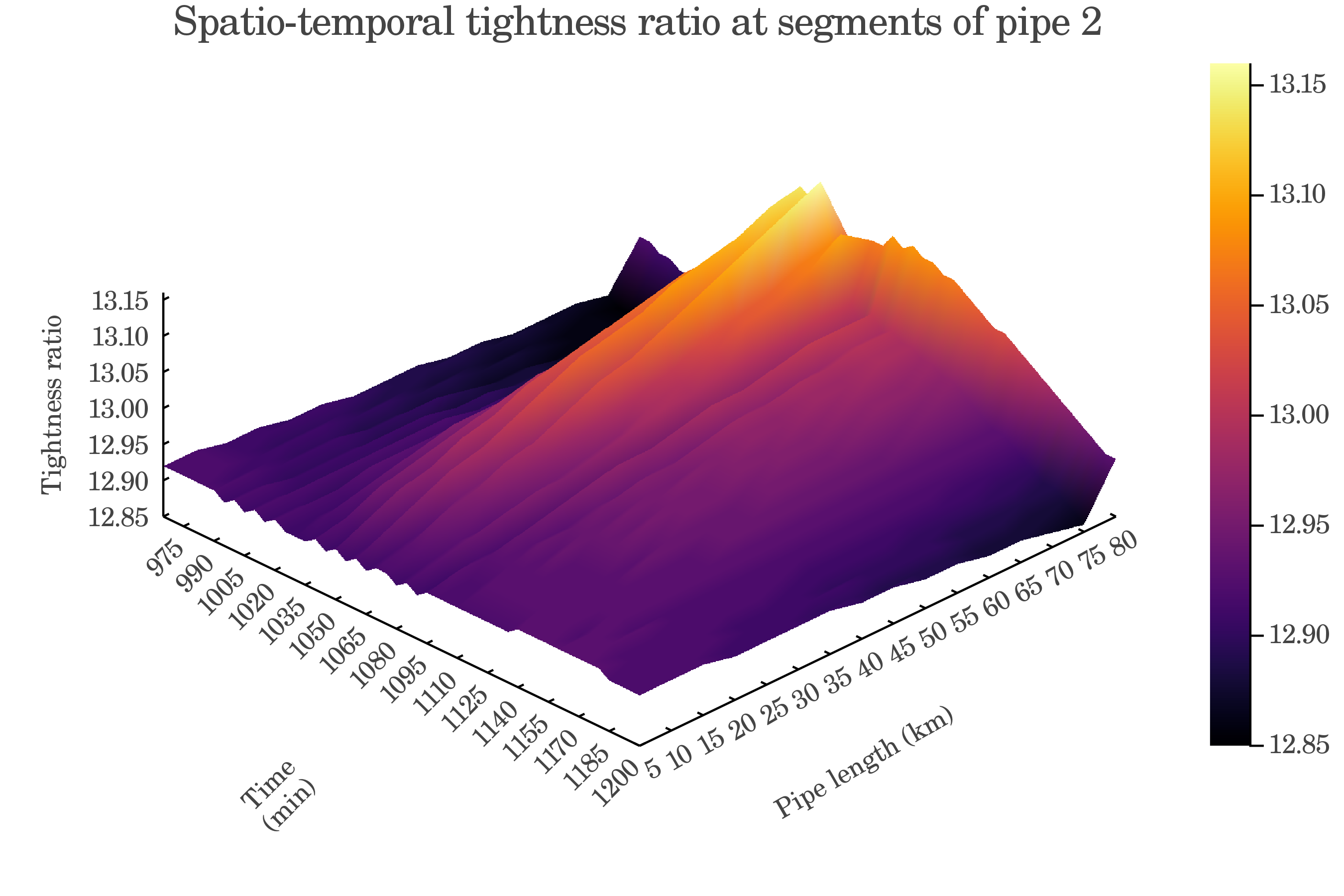}
    \vspace{-0.95cm}
    \caption{Tightness of eigenvalues for SDP matrix for pipe $2$}
    \label{fig:3d_eigenvalue}
\end{figure}

\par
\begin{figure}[h!]
    \centering
    \vspace{-0.35cm}
    \includegraphics[width=.9\linewidth]{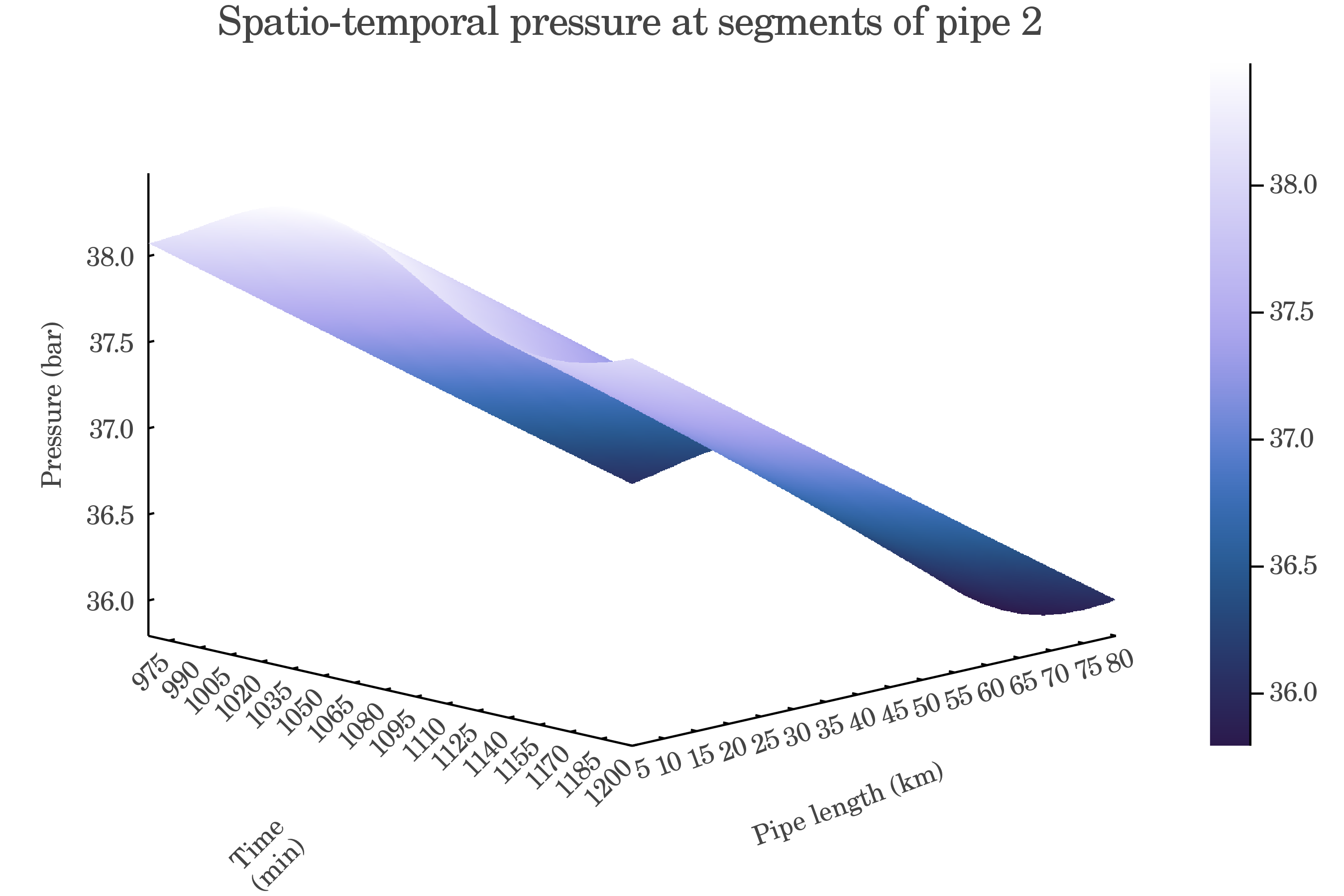}
    \vspace{-0.35cm}
    \caption{Pressure across pipe 2 at different time steps}
    \label{fig:3d_pressure}
    \vspace{0.35cm}
    \includegraphics[width=.9\linewidth]{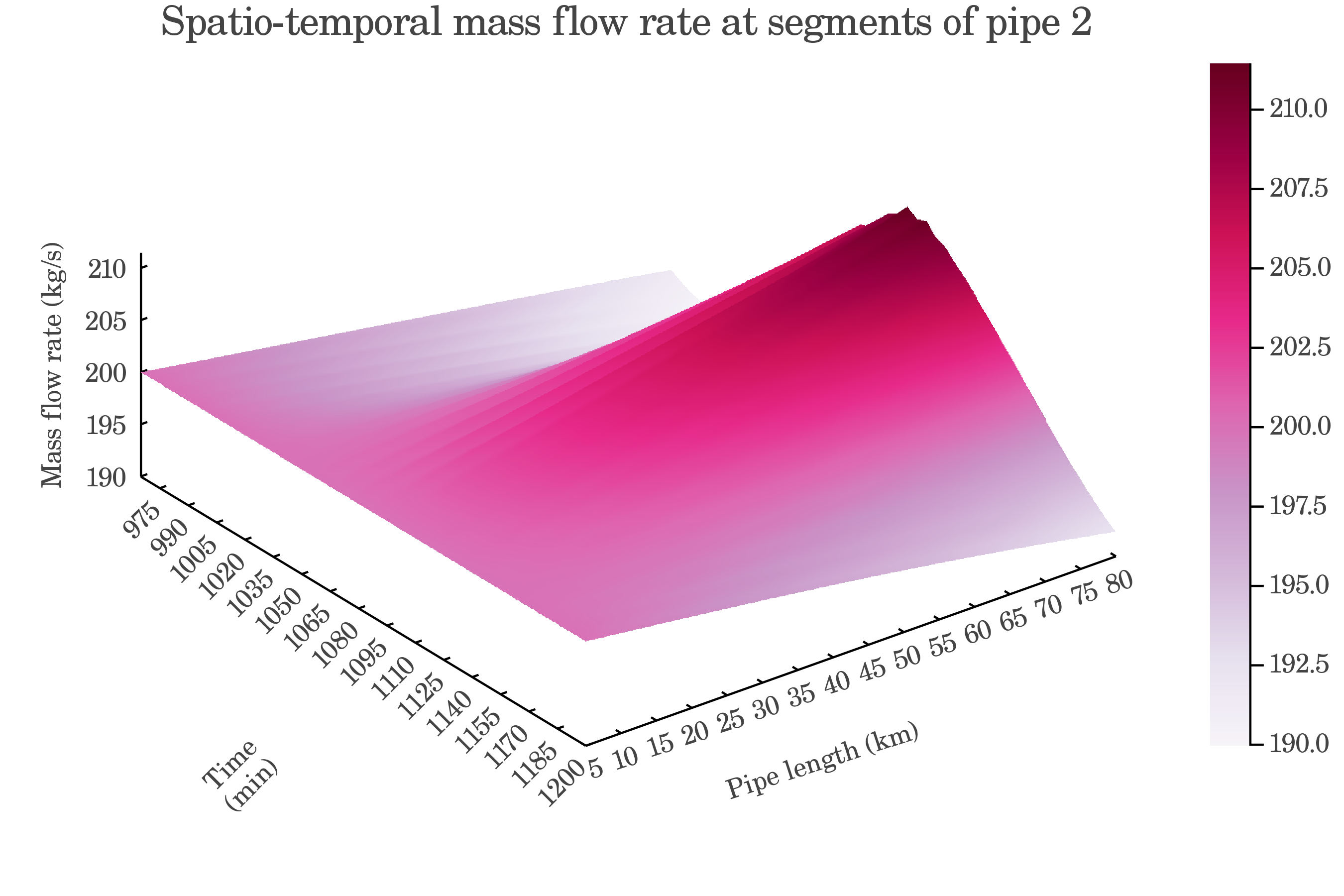}
    \vspace{-.35cm}
    \caption{Mass flow rate across pipe 2 at different time steps}
    \label{fig:3d_mass}
    \vspace{-0.35cm}
\end{figure}

 According to the reported tightness ratio values, the procured answer with our proposed method is tight and physically meaningful for all pipe segments during all time intervals of the day. Fig. \ref{fig:3d_eigenvalue} provides the values for the tightness of the solution obtained with the proposed model across pipe 2. Using smaller time intervals allows us to investigate the characteristics of each pipe of the system in both spatio-temporal scopes. This results in an exhaustive understanding of the natural gas dynamics that are taking place inside the pipes at all locations and times. The visualization of the natural gas mass flow throughout pipelines can  convey a better perception of the \textit{linepack} at all times. 
Figs. \ref{fig:3d_pressure} and \ref{fig:3d_mass} illustrate the pressure and mass flow rate {across the} segments of pipe 2, respectively. Fig. \ref{fig:3d_pressure} shows how pressure is adjusted throughout the day to feed the load. It is observed that inside a pipe, pressure decreases gradually in the direction of gas flow, which in this case is from the $1^{st}$ segment of the pipe toward its last segment. As exhibited in Fig. \ref{fig:3d_mass}, the mass flow rate at the end of pipe 2 displays the same trend as the demand curve. At the beginning of this pipe, the mass flow rate is almost uniform because it stems from a natural gas source. \\

\par
\begin{figure}[h!]
    \centering
    \vspace{-0.15cm}
    \includegraphics[width=.9\linewidth]{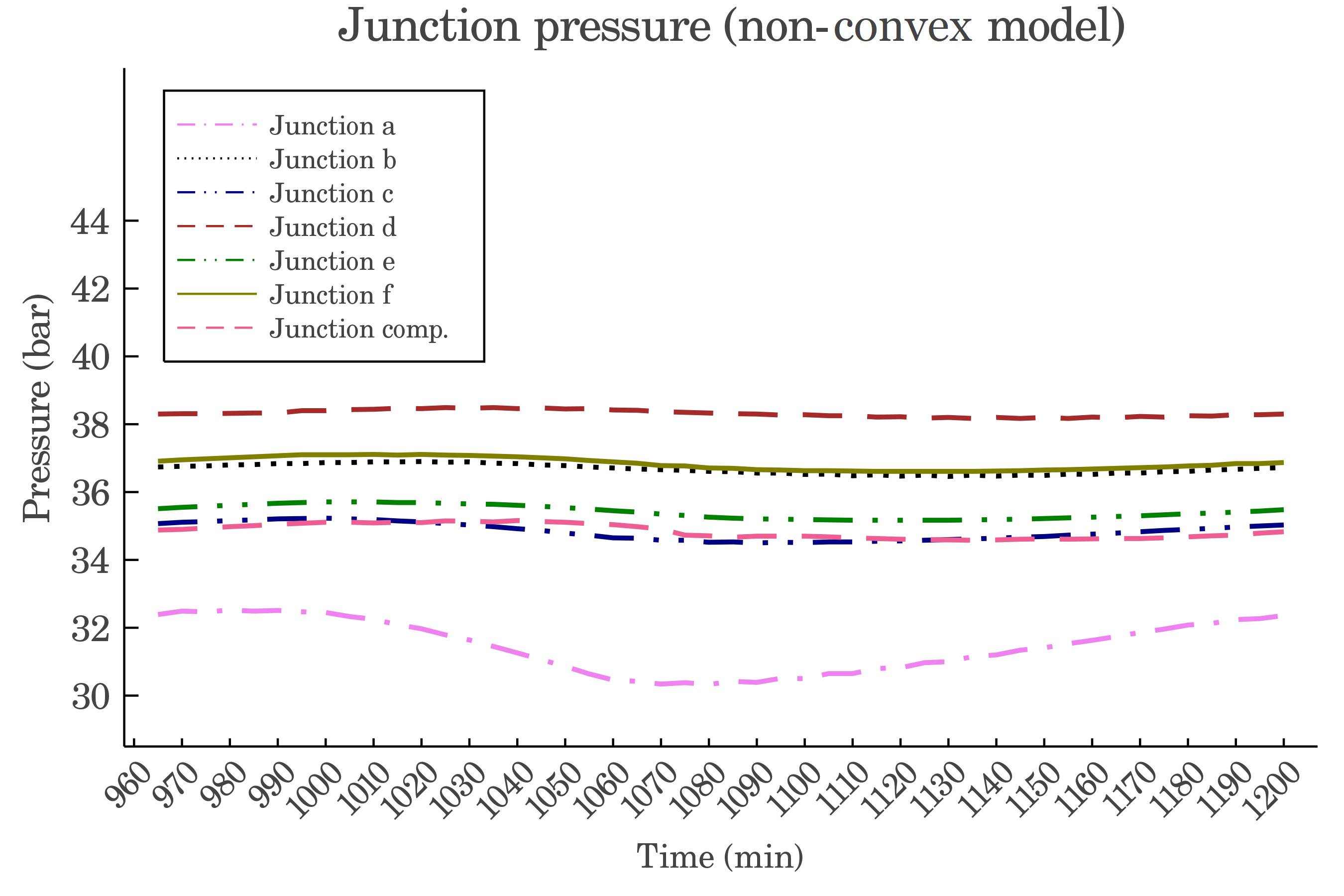}
    \vspace{-0.35cm}
    \caption{Pressures at the 6-junction network acquired with the dynamics model}
    \label{fig:dynamics_pressure}
    \vspace{0.35cm}
    \includegraphics[width=.9\linewidth]{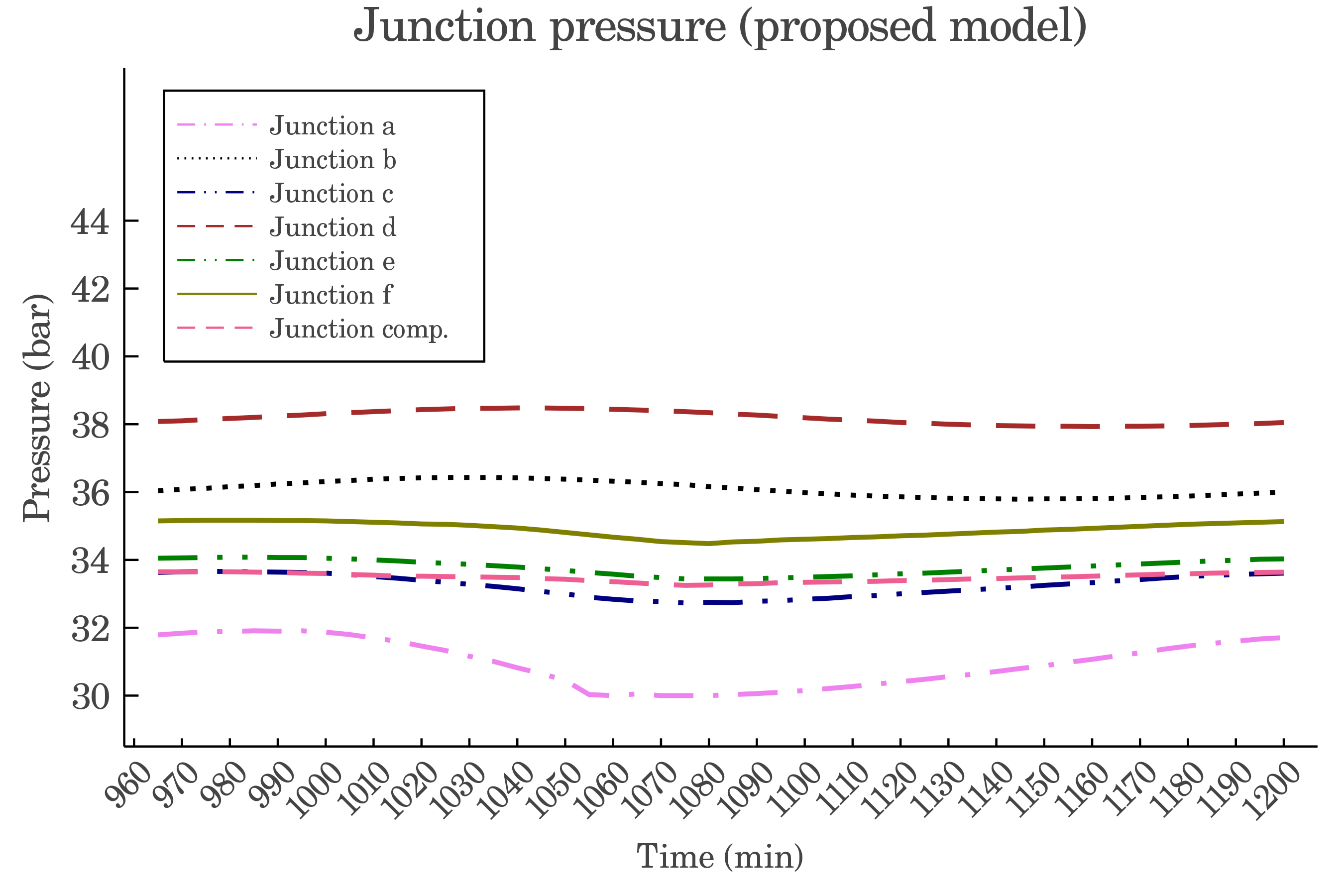}
    \vspace{-0.35cm}
    \caption{Pressures at the 6-junction network acquired with the proposed model}
    \label{fig:relaxed_pressure}
\end{figure}

 As shown in Table \ref{table:validation}, it is also noticed that the objective value of the results procured by implementing the proposed model, while the rendered solution is feasible for the original non-convex problem, is smaller than the objective value obtained by applying the non-linear and non-convex model. While the solution procured by the non-convex model might be a local solution, the proposed convex model can render a higher quality solution with a lower operating cost which is also feasible for the original non-convex problem.
Figs. \ref{fig:dynamics_pressure} and \ref{fig:relaxed_pressure} show the pressures at junctions of the 6-junction natural gas network acquired by the non-convex and the proposed model. It is noticed that except for junction `f', the pressures {at} all junctions reached roughly equal patterns and values for these two models. Even for junction `f', the pattern in both of these figures is the same during the simulation horizon. When looking more closely, it is observed that the pressure values for the proposed model are slightly less than the pressure values obtained by the non-convex model. The governing equations of the OGF problem are more sensitive to pressure difference than the pressure value itself. The values and figures presented in this section strongly suggest that the presented convex relaxation method with the rank minimization problem addresses the shortcomings of the non-linear models. It will deliver a better quality solution with a smaller computation burden and an optimal solution that is tight enough to be feasible for the original non-convex problem.

\par The model choice and the relaxation results of the proposed work display considerably better performance compared with past works in the literature due to superiority in model accuracy, granularity, and relaxation method tightness. Table \ref{tab:Rev_2_compare_model} shows a comparison of 4 model choices and displays the percentage error values of their results against the PDE model. The listed works use gas flow models with simplifying assumptions (such as neglecting acceleration) which result in a formulation much similar to Weymouth model. It is observed that {the} model choice and discretization level of the presented work lead to {a} substantial reduction in modeling error and relaxation error values. The quality of the decisions of other models and their convex-relaxed forms are not comparable with the PDE results, and their relaxations lead to small tightness ratios.

\begin{table}[h!] \footnotesize \centering
	\caption { Performance comparison of different models and granularity choices } \label{tab:Rev_2_compare_model}
\resizebox{\columnwidth}{!}{ 
\begin{tabular}{ccccc}\hline\hline 
Reference  & \makecell{Spatio-temporal \\ discretization} & \makecell{Non-convex \\model error} & \makecell{Tightness \\ ratio}\\ \hline
\cite{fan2020multi} & $\Delta x = 80 km, \;\;\Delta t = 1 hr$ & 17.7\% &   3.23 \\
\cite{Chen2018}  &  $\Delta x = 20 km, \;\;\Delta t = 1hr$  & 14.4\% &  2.07 \\ 
\cite{yang2017effect}  &  $\Delta x = 20 km, \;\;\Delta t = 15 min$ & 4.0\% &  NA (2.05 with SOC) \\
Proposed  &  $\Delta x = 5 km, \;\;\Delta t = 5 min$ & 1.1\% &  2.43 \\
\hline\hline
\end{tabular}}
\end{table} 

\begin{figure}[h!]
    \centering
 \includegraphics[width=7.5cm]{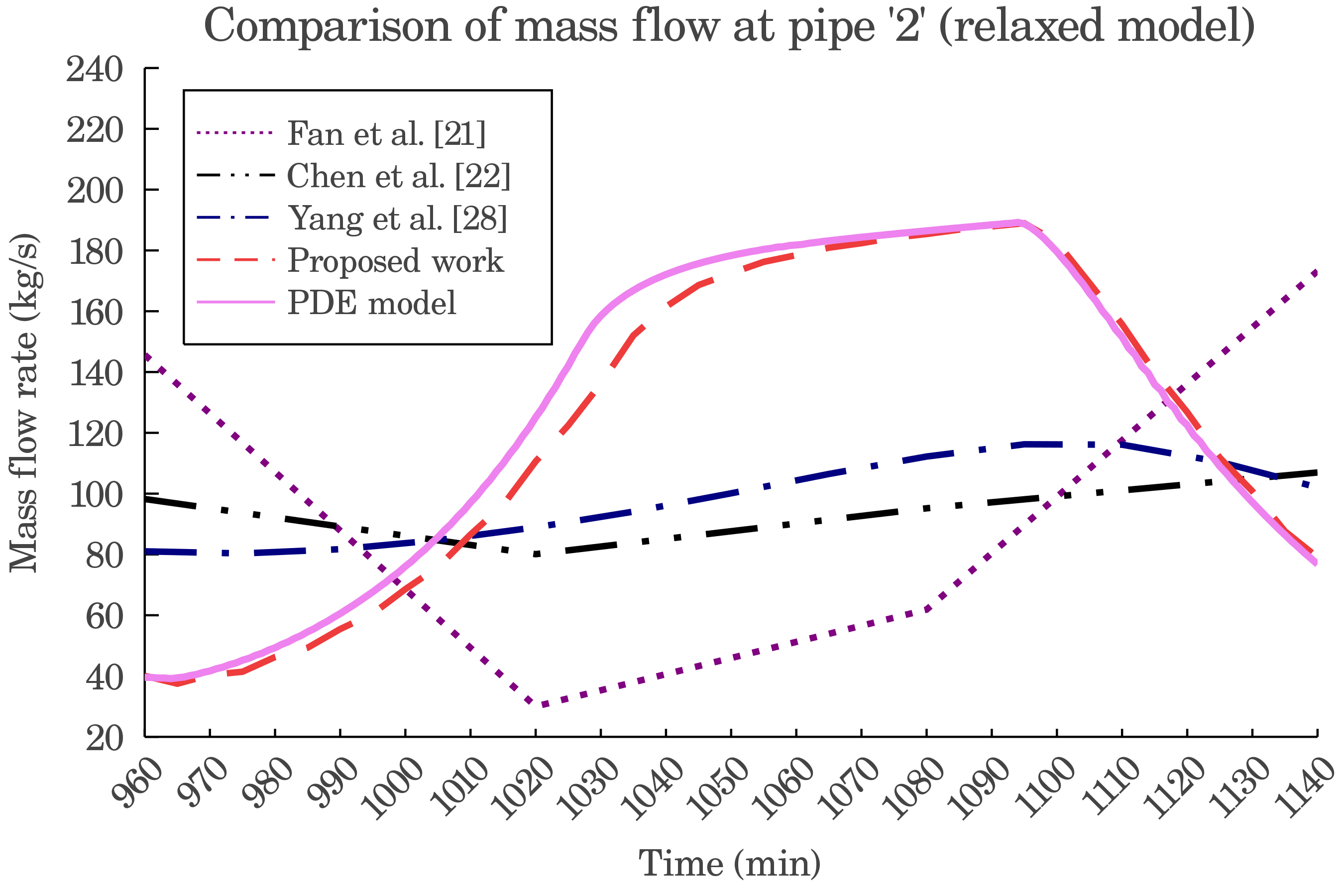}
  \vspace{-.45cm}
  \caption{ Comparison of mass flow rate at pipe 2 when applying relaxed models}
  \label{fig:resp2_compare_mass_relax}
\end{figure}

Fig. \ref{fig:resp2_compare_mass_relax} visualizes a comparison {of} the results of relaxed models proposed in \cite{fan2020multi}, \cite{Chen2018}, \cite{yang2017effect}, and this work. Although {the} authors in \cite{yang2017effect} did not propose any relaxation scheme, {the} SOC method was applied to their non-convex model to present a better comparison. Contrary to the other methods, the results of the proposed relaxation scheme are both consistent with the results of the PDE model and the proposed model's non-convex original form.

\vspace{-.35cm}
\subsection{A Larger Interdependent Electricity and Natural Gas Network}
\begin{figure}[b!]
	\vspace{-0.4cm}
    \centering
    \includegraphics[width=.9\linewidth]{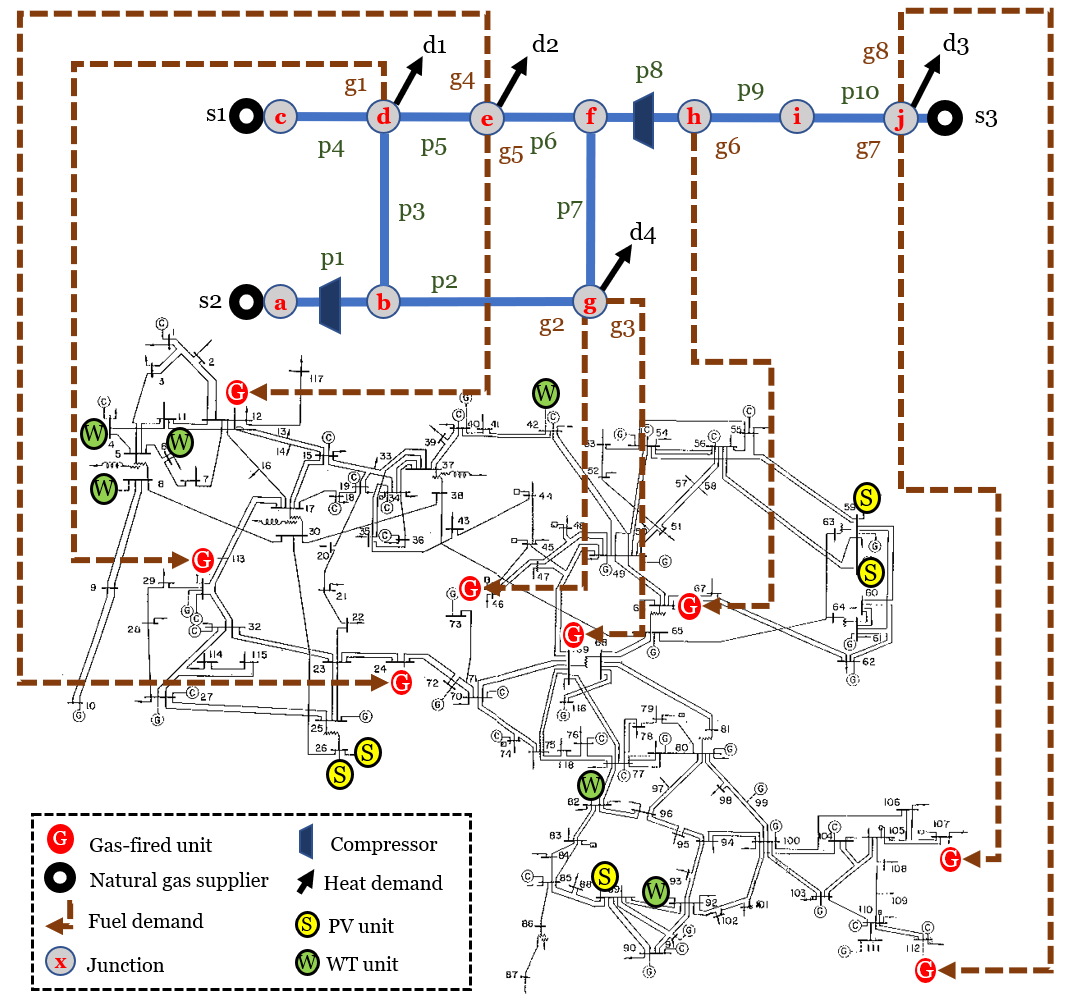}
    \vspace{-0.45cm}
    \caption{Structure of the large network case}
      \vspace{-0.25cm}
  \label{fig:118_10}
\end{figure}

The proposed model is also capable of dealing with larger  problems. In this part, the simulation focused on a 4-hour period operation problem of a  larger  interdependent network to illustrate the scalability of the proposed method. The composition of the modified IEEE 118 bus network with a 10-junction natural gas network is considered for this case. It is supposed that the modified network is highly renewable integrated. The time-step {length} for {the}  simulation of the power network is 1 hour, while the time intervals {for} the natural gas network are equal to 5 minutes. The 10-junction 10-pipe natural gas network has a mesh structure, and two compressors with the maximum compression ratio of 1.1 are placed on pipes 1 and 8. It serves two types of loads, the heat demand and the fuel demand for gas-fired units of the electricity network. The structure of the coupled network is displayed in Fig. \ref{fig:118_10}. Overall, 8 gas-fired units are present in the power system, and the results for their hourly dispatch are displayed in Table \ref{table:dispatch_118}.
\begin{table}[h!]
\setlength\tabcolsep{4.5pt} 
	\footnotesize \centering
	\vspace{-0.55cm}
	\caption {Gas-fired unit hourly dispatch in 118 bus system}
    \begin{tabular}{ccccccccc} \hline\hline
        \textbf{Unit ID} & 1 & 2 & 3 & 4 & 5 & 6 & 7 & 8\\ \hline\hline
        \textbf{Hour 14 (MW)} & 0 & 0 & 0 & 0 & 0 & 0 & 0 & 35\\ \hline
        \textbf{Hour 15 (MW)} & 0 & 0 & 0 & 0 & 75 & 149 & 100 & 100\\ \hline
        \textbf{Hour 16 (MW)} & 0 & 0 & 57 & 100 & 550 & 185 & 100 & 100\\ \hline
        \textbf{Hour 17 (MW)} & 100 & 100 & 100 & 100 & 550 & 185 & 100 & 150\\ \hline\hline \label{table:dispatch_118}
    \end{tabular}
    \vspace{-0.55cm}
\end{table}
\par Similar to the previous case, applying the proposed relaxation method to a more extensive problem also leads to the procurement of very tight solutions. The mean values for the tightness ratios over all segments of each pipe for all time steps are reported in Table \ref{table:tightness_118}. According to these values, it can be argued that the procured solution is tight and reliable for this case. The overall average tightness ratio of the system is equal to 8.9, which is satisfactory.
\begin{table}[h!]
    \vspace{-0.15cm}
	\footnotesize \centering
\setlength\tabcolsep{4.5pt} 
	\caption {Average spatio-temporal tightness ratio of pipes}
    \begin{tabular}{ccccccccccc} \hline\hline
        \textbf{Pipe} & 1 & 2 & 3 & 4 & 5 & 6 & 7 & 8 & 9 & 10 \\ \hline
        \textbf{Tightness} & 9.7 & 7.9 & 8.3 & 9.8 & 9.6 & 8.6 & 5.1 & 9.3 & 10.0 & 9.3 \\\hline\hline \label{table:tightness_118} \vspace{-0.55cm}
    \end{tabular}
\end{table}
\par
\begin{figure}[h!]
	\vspace{-0.1cm}
    \centering    \includegraphics[width=1.0\linewidth]{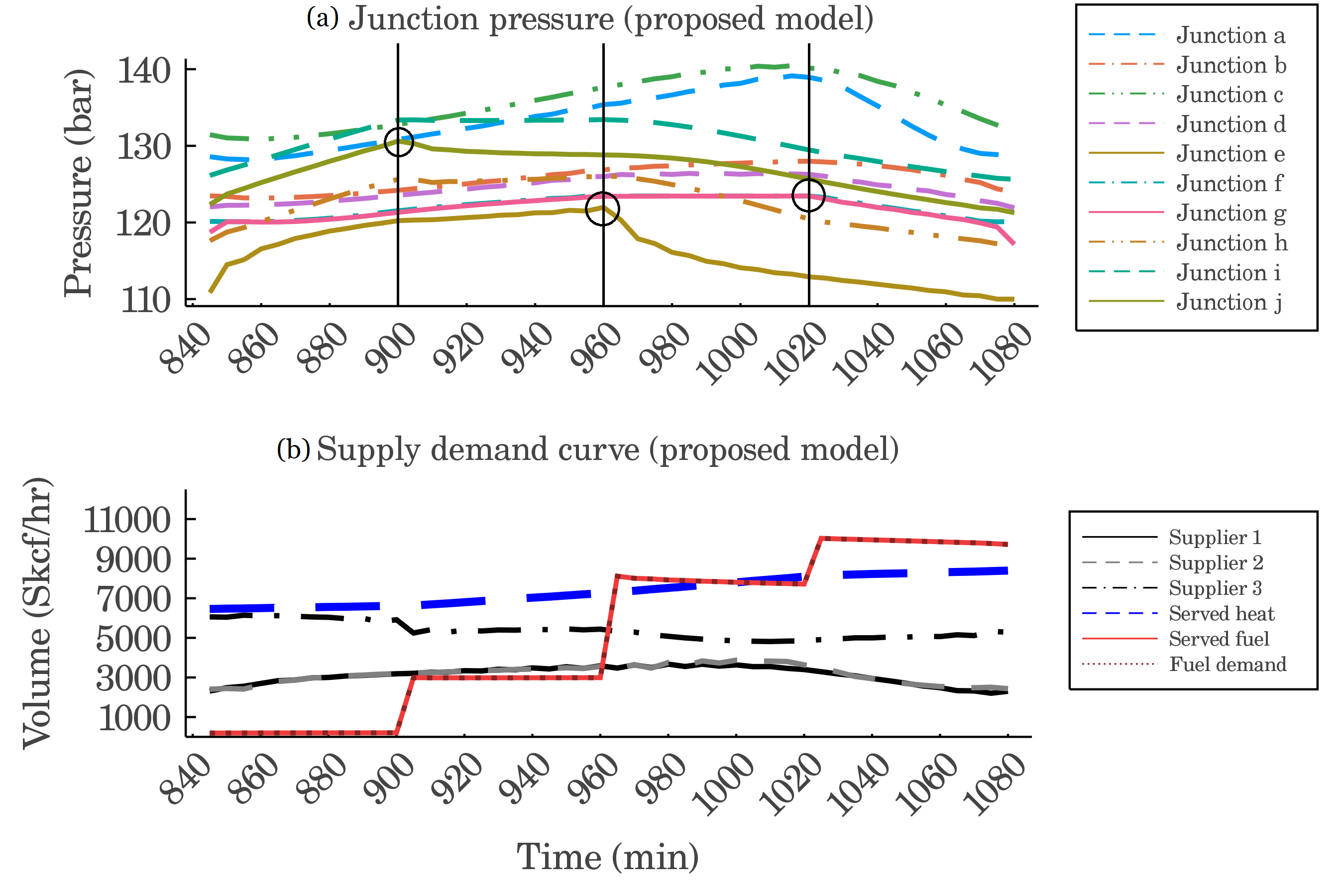}
    \vspace{-0.75cm}
    \caption{(a) Pressure and (b) supply-demand curve in the large case}
    \label{fig:118_supply_pressure}
\end{figure}

 By comparing the supply-demand curve and the junction pressure curve of this case, it can be observed that the volume of suppliers will not change over time to serve the demand. It is fascinating to notice that the adjustments {in} the pipeline pressures serve the variations in the natural gas demand  rather than changes in the supplier gas volume. In Fig. \ref{fig:118_supply_pressure} (a) and (b), respectively, the pressure at the junctions and the supply-demand curve of the natural gas system are displayed. It is observed that for this setting, all the fuel demand is served. Also, the hourly spikes in the fuel demand curve are consistent with the generation dispatch of units given in Table \ref{table:dispatch_118}. 

\par Unlike electricity, natural gas does not travel instantly, and the dynamics inside pipelines are not observed momentarily. Therefore, the total demand of this system is not necessarily equal to its total supply at each moment. More precisely, it is the changes in the pressure {at} the junctions which is an indicator of the mass within pipelines. 
To meet the variations in demand, the natural gas pressure at the demand junction plays the most crucial role. At min 900 (hour 15), units 5, 6, and 7 start to come online and join the previously committed unit 8. The pressure {at} junction `j' at this moment drops to meet the increased demand placed on this junction. A similar story happens with the pressure at other demand junctions at the following two-hourly dispatch marks, when units 3 and 4 come online at min 960 (hour 16), and when units 1 and 2 come online at min 1020 (hour 17), pressures at junctions. The impact of additional fuel demand at these two hours can be observed on the pressure fall at minutes 960 and 1020 for the junctions on which these generators are placed, as shown in Fig. \ref{fig:118_10}, i.e., junctions `e' and `g'.

\section{Conclusion}
Inside a natural gas network with a varying natural gas demand, the output of gas suppliers does not necessarily follow the demand trend. This paper demonstrates how the varying natural gas mass demand is matched through alterations in pressure values.
The incapability of {the} Weymouth equation in modeling the demand variations is illustrated. Proper modeling of natural gas dynamics is crucial for the short-term operation of the electricity grid in light of fuel security concerns. 
The short-term operation problem requires utilizing an OGF model that is both accurate and exact. A rank minimization method is proposed for the relaxed OGF model with dynamics to address this challenge.
Simulation results of applying the presented approach indicate that it can successfully incorporate the dynamics of the natural gas network while significantly reducing the computation time compared to the non-convex method. Due to its extremely high tightness ratio, the solution procured with the proposed method is feasible for the original non-convex problem with superior quality than the non-convex model. Results of the proposed model are also compared with other works in the literature. The observations clarify the importance of model choice and granularity on {the} accuracy of the results. The scalability of the proposed approach to efficiently solve larger problems is also demonstrated. The resulting convex transient OGF model in this work can be viewed as an efficient tool which can be utilized in the coordinated short-term operation problem of the electricity grid and natural gas network with dynamics.

\bibliographystyle{IEEEtran}
\bibliography{short_term}

\vspace{-1cm}
\begin{IEEEbiographynophoto}{Reza Bayani}
(S'20) received the B.Sc. degree in Electrical Engineering from Sharif University of Technology, Tehran, Iran, in 2013 and the M.Sc. degree in Electrical Engineering from Iran University of Science and Technology, Tehran, Iran, in 2016. He is currently pursuing his Ph.D. degree in Electrical and Computer Engineering at University of California San Diego, La Jolla, USA, and San Diego State University, San Diego, USA. His current research interests include power system operation and planning, machine learning, and electric vehicles. He is a Ph.D. fellow of California State University's Chancellor's Doctoral Incentive Program.
\end{IEEEbiographynophoto}

\vspace{-1cm}
\begin{IEEEbiographynophoto}{Saeed D. Manshadi}
(M'18) is an Assistant Professor with the Department of Electrical and Computer Engineering at San Diego State University. He was a postdoctoral fellow at the Department of Electrical and Computer Engineering at the University of California, Riverside. He received the Ph.D. degree from Southern Methodist University, Dallas, TX; the M.S. degree from the University at Buffalo, the State University of New York (SUNY), Buffalo, NY and the B.S. degree from the University of Tehran, Tehran, Iran all in electrical engineering. He serves as an editor for IEEE Transactions on Vehicular Technology. His current research interests include smart grid, transportation electrification, microgrids, integrating renewable and distributed resources, and power system operation and planning.
\end{IEEEbiographynophoto}
\end{document}